\documentclass[preprint2]{aastex}
\usepackage[dvips]{epsfig}

\shorttitle{Mass Models}
\shortauthors{Barnes, Sellwood, \& Kosowsky}

\slugcomment{resubmitted to AJ}

\newcommand{\chisq}{\ensuremath{\chi^2_r}}

\newcommand{\biwt}{\ensuremath{\chi^2_b}}
\newcommand{\biwtm}{\ensuremath{\chi^2_{b,\rm min}}}
\newcommand{\ignore}[1]{\relax}

\newcommand{\ml}{\ensuremath{\Upsilon}}
\newcommand{\mli}{\ensuremath{\Upsilon_I}}
\newcommand{\mld}{\ensuremath{\Upsilon_{D,I}}}
\newcommand{\mlb}{\ensuremath{\Upsilon_{B,I}}}

\newcommand{\eg}{{\it e.g.}}
\newcommand{\ie}{{\it i.e.}}

\begin{document}
\title{Mass Models for Spiral Galaxies from 2-D Velocity Maps}
\author{Eric I. Barnes, J. A. Sellwood, \& Arthur Kosowsky}
\affil{Department of Physics \& Astronomy, Rutgers 
University, Piscataway, NJ 08854}
\email{barnesy@physics.rutgers.edu}
\email{sellwood@physics.rutgers.edu}
\email{kosowsky@physics.rutgers.edu}

\begin{abstract}

We model the mass distributions of 40 high surface brightness spiral
galaxies inside their optical radii, deriving parameters of mass
models by matching the predicted velocities to observed velocity maps.
We use constant mass-to-light disk and bulge models, and we have tried
fits with no halo and with three different halo density profiles.  The
data require a halo in most, but not all, cases, while in others the
best fit occurs with negligible mass in the luminous component, which
we regard as unphysical.  All three adopted halo profiles lead to fits
of about the same quality, and our data therefore do not constrain the
functional form of the halo profile.  The halo parameters display
large degeneracies for two of the three adopted halo functions, but
the separate luminous and dark masses are better constrained.
However, the fitted disk and halo masses vary substantially between
the adopted halo models, indicating that even high quality 2-D optical
velocity maps do not provide significant constraints on the dark
matter content of a galaxy.  We demonstrate that data from longslit
observations are likely to provide still weaker constraints.  We
conclude that additional information is needed in order to
constrain the separate disk and halo masses in a galaxy.
\end{abstract}


\section{Introduction}

The mass of a galactic stellar population has proved to be a
particularly difficult quantity to determine.  In principle, an
accurate value would place a constraint on the star formation history,
stellar mass function, and metallicity.  But it is perhaps more
important for dynamics because it affects estimates of the dark matter
content of the inner parts of galaxies.  There are two possible
approaches to estimating the stellar mass: dynamical measurements and comparison
of the observed broad-band colors or spectra with theoretical predictions
from population synthesis.

Theoretical models to synthesize a stellar population
\citep[\eg,][]{ay86,bc93,wor94,bell03} use established stellar models
and assumptions about the likely mass function, star formation
history, etc.\ to predict the luminosity in a broad-band color.  The
expected luminosity in almost any color band is insensitive to the
low-mass slope of the adopted mass function, since large proportions
of low-mass stars can be added or removed with little effect on the
total luminosity.  The resulting mass-to-light values, \ml, are
thought to be reliable to within a factor of 2 \citep{bdj01}.

The dynamical approach is scarcely more successful.  The typical
circular flow pattern of gas in a spiral galaxy affords a rather
precise estimate of the overall mass profile within the radial range
explored by the data.  But it is much harder to determine the fraction
of the enclosed mass that comes from the starlight, since an unknown
fraction of dark matter is also expected to contribute.  A common simplifying
assumption is that the disk mass profile traces that of the disk light with
a constant, but unknown, \ml.  But \citet{va85} and \citet{lf89} stress that,
for an axisymmetric flow pattern, the disk and halo masses are almost 
completely degenerate, even for galaxies with the best
available rotation curve data from neutral hydrogen observations.

An additional factor that contributes to this difficulty is the
unknown density profile of the dark matter halo.  Previous studies
have tried a number of functional forms, some motivated by
cosmological simulations, but with no decisive conclusions
\citep[\eg,][]{nav98,voj02}.  It is scarcely surprising that very weak
constraints result from fitting a 1-D rotation curve as the sum of a
contribution from the luminous matter, with an unknown mass scaling,
plus a contribution from dark matter with both an unknown functional
form and an unknown density normalization.

\citet{wsw01b}, \citet{weiner03}, and \citet{ksr03} were able to obtain 
more precise estimates (within $\sim 10\%$, Weiner et al.\ 2001b) of the 
stellar disk mass through modeling the non-axisymmetric flow pattern in disk
galaxies.  These were time-consuming studies, requiring a large grid
of hydrodynamical simulations over the two unknowns, \ml\ and the
pattern speed of the bar or spiral.  Alternatively, \citet{bott97}
measures the velocity dispersion of the stars in the disk, to obtain a
disk mass estimate \citep[see also][]{verh03}.  While these more
challenging methods may yet be needed to break the disk-halo mass
degeneracy, we here examine the question of whether fitting simple
axisymmetric models to high-quality, 2-D optical velocity maps offers
any tighter constraints over previous attempts at fitting 1-D rotation
curves from longslit data or lower-resolution 2-D maps from aperture
synthesis.

Velocity maps at low spatial resolution have long been available from
neutral hydrogen aperture synthesis observations
\citep[\eg,][]{bos77}.  More recently, velocity maps have become
available at very high spatial resolution from mm-wave observations of
molecular gas \citep{jogee02,sofue03,simon03}.  Most optical data has
continued to be obtained using longslit spectra
\citep[\eg,][]{cour97,sr01,vogt04}, but a few maps are available from
integral field units; \citet{corr92} and \citet{pw00} used
Fabry-Perot maps, \citet{and02} have used fiber bundles, and the lens
array spectrograph SAURON has been used to investigate early type
galaxies \citep{davies01}.

Palunas \& Williams present the largest sample of galaxies with
homogeneously gathered Fabry-Perot maps and $I$-band photometry
available so far.  They showed that the inner rotation curves
could be fitted by constant \mli\ mass models and no dark matter, but
with a rather wide range of \mli\ values, when mass discrepancies in
the outer parts were excluded from the fit.  We here present a 
re-analysis of these galaxies to
determine what can be concluded about the likely dark matter content
of galaxies from fitting axisymmetric models to these data.

Our objective here is to try to determine whether our data place meaningful
constraints on the dark matter content of the galaxies in our sample
and whether one halo profile
or another yields a sigificantly better fit within the radial range spanned 
by our data.  These objectives are quite different from 
testing theoretical predictions from some model of galaxy formation.

We find that 2-D velocity maps do provide some advantage over 1-D
rotation curves, in that they reduce the statistical uncertainties in estimated
parameters.  However, the disk mass, and therefore also the dark
matter content of a galaxy, is still strongly dependent on the adopted
halo density profile.  We support these conclusions by fitting mass
models to both the 2-D velocity map (\S\ref{dmfits}) and to a 1-D
rotation curve (\S\ref{longslit}) which we derive from data along only
the major axis of our velocity maps.

\section{Data}\label{data}

The data used in this paper were kindly provided by Povilas Palunas
and Ted Williams (see Palunas \& Williams 2000, hereafter PW, for a
complete description of the data and reductions).  We have $I$-band
images and Fabry-Perot (hereafter FP) velocity maps for 74 high
surface-brightness spiral galaxies.  PW report the average seeing to
be $\approx 1.5\arcsec$.  We adopt the magnitudes, luminosities, and
distances reported by PW; distances are simply Hubble distances
assuming $H_0=75 \: \mbox{km s}^{-1} \: \mbox{Mpc}^{-1}$.

Galaxies in this sample were selected to lie in a particular part of
the sky, to be of later Hubble type, to not be classifed as barred in
the RC3 catalog \citep{dev91} (although many are, in fact, barred), to
not be obviously disturbed, to have a suitable angular size for the FP
instrument, and to have redshifts in the range 4000 -- 5000 km/s.  The
data were taken to obtain optical Tully-Fisher distances of galaxies
to map departures from the Hubble flow in the general direction of the
Great Attractor, which may have resulted in a somewhat larger than
normal fraction of galaxies in clusters, although roughly half (35 of
74) are in the field.  See \S~4.1 for a discussion of the impact of
the Great Attractor on the distances adopted for these galaxies.
Apart from this, the sample seems unlikely to have any sinister
biases relevant to their mass distributions.

We have created mass models for a subsample of 40 galaxies.  The
T-types of this subsample (from the RC3) range between 2 (Sab) and 10
(Irr) with most being 4-6 (Sbc-Sc).  The radii at which the surface
brightness is 23.5 mag/arcsec$^2$ $R_{23.5}$ range from 24 to
104\arcsec.  Their average luminosity is $3\times10^{10} \:L_{\sun}$
and ranges between $8\times10^9$ and $8\times10^{10} \:L_{\sun}$.  The
average I-band central surface brightness is 19.6 $\mu$, with the
brightest galaxy at 18.3 $\mu$ and the faintest at 21.3 $\mu$.
Maximum circular speeds for these galaxies range from approximately
100 to 300 km/s with an average of 170 km/s.  The 34 galaxies in their
sample which we have omitted from this study are listed in
Table~\ref{omitted}.  The majority were discarded because they are
close to edge-on, having inclinations greater than $75^{\circ}$, which
raises a number of difficulties: (1) The finite thickness of their
disks becomes a significant fraction of the projected minor-axis and
invalidates our simple razor-thin disk model (see \S\ref{decomp}).
(2) Extinction problems become severe, and dust corrections large and
uncertain.  (3) The velocity map is less well resolved, and gas
emission on the minor axis can more easily skew the estimated velocity
from the major axis, which could lead to an estimated rotation curve
that may be more slowly rising than is really the case.  A few
galaxies were discarded for other reasons, such as a lack of kinematic
data in their central regions.

We did not discard galaxies on the basis of their morphology and have
persisted in fitting axisymmetric models to a few that manifestly have
strong bars.  The reason for this is that almost all galaxies in the
sample have weak non-axisymmetric features, and judgements of which
galaxies to discard would have been arbitrary and subjective.
Modeling of the non-axisymmetric flow pattern would be superior, but
it is a major task for even one galaxy \citep{wsw01b,ksr03} which we do
not attempt here.

Throughout this paper, we fix the position angle $\phi_0$,
inclination angle $i$, and systemic velocity $v_s$ of each galaxy at
the values deduced from earlier fits of axisymmetric flow
patterns to our kinematic data \citep{bs03}.  Briefly, we found the
values of these parameters that minimized $\chi^2$ between the 2-D
velocity map from the FP data and a planar axisymmetric flow pattern.
The fit therefore also yields a set of mean orbital speeds, and their
uncertainties, at equally-spaced radii which we show as the ``data''
in subsequent figures.  Since we fit axisymmetric mass models, we
could, in principle, fit this 1-D rotation curve, which is basically
the speed averaged in annular bins, provided we compute the weights
correctly.  However, our use of the biweight (\S3.3) requires a
fit to the full 2-D velocity map.

The use of kinematically-derived projection parameters is possible
only because we have velocity maps.  Rotation curves derived from
long-slit observations must adopt inclination and major-axis position
angles from the brightness distribution in the galaxy.  We discuss
how much this might affect the results in \S7.

\section{Methods}

We build a mass model for each galaxy in our sample based on the
measured light distribution in the $I$-band and a possible dark matter
component, and then compare the 2-D velocity map expected from gas on
circular orbits in our model with that observed from the H$\alpha$
emission.

We assume separate constant $I$-band mass-to-light, \mli, for the disk
and bulge components and neglect the possibility of \mli\ gradients
within each component.  This is the usual assumption, despite
established color gradients in galaxy disks \citep[\eg,][]{dj96}.  We
justify this assumption because the unknown correction for \mli\
gradients will have a small effect, as shown by \citet{wsw01a}: the
rotation curve shape for the disk is largely determined by the mass
distribution in the dense inner parts, and changes to the density
gradient at larger radii have a minor effect.

We also neglect any contribution from gas.  In previous models where
HI maps are available \citep[\eg,][]{bro92,verh97}, the contribution
from atomic gas to the rotation curve is generally very small in the
inner parts, but rises toward the optical edge -- to $\lesssim 10\%$
of the central attraction in the most extreme case.  Molecular gas is
generally more concentrated towards the inner parts
\citep[\eg,][]{reg01} where its contribution may be somewhat larger,
but we could not find any information on the molecular gas content of
our galaxies.  HI fluxes for 28 galaxies in our sample are available
from NED.\footnote{The NASA/IPAC Extragalactic Database (NED) is
operated by the Jet Propulsion Laboratory, California Institute of
Technology, under contract with the National Aeronautics and Space
Administration.}  Since we find gas mass fractions typical of the
galaxies in these other studies, neglect of the gas contribution
should result in only a slight systematic overestimate of the \mli\
for the stellar component, unless the dark matter fraction is large.

We consider three different radial density profiles for the dark
matter halo, which we assume to be spherical for simplicity.  Two of
our halo functions are standard: the pseudo-isothermal model with a
constant-density core and a model with a broken power law; the third
is a simple, scale-free power-law.  Our
objective is not to test any particular model of dark matter
halo formation, but to determine how well our data constrain the halo
contribution to the rotation curve in the disk plane.  It is possible
that halos are spheroidal, or even triaxial; the unknown shape of the
halo may alter the inferred mean density profile slightly, but our
data cannot constrain such subtleties.  The calculation of the
contribution to the circular velocity from a spherical halo model is
straightforward.

\subsection{Bulge-Disk Decomposition}\label{decomp}

Since we wish to allow the disk and the bulge to have separate \mli\
values, we must separate the light distribution into these two
components.  While we do this with care, the central attraction is not
strongly sensitive to the precise decomposition, because we ensure
that the \mli\ of the two components in our models do not differ by a
large factor.  (We regard it as unlikely on physical grounds that $\mlb
\ll \mld$; we discard all such solutions that emerge from an
unconstrained minimization, often with $\mlb\ \ll 0.1$, and repeat
the fit with the constraint that $\mlb = \mld$.)  We also ensure that
all the light is included in one component or the other.  Therefore
the total luminous mass profile remains nearly the same no matter how
the light is divided.  The central attraction depends slightly on
whether the mass is in a flat disk or a spheroid, with a maximum
possible difference in circular speeds of $\sim 15\%$ (Binney \&
Tremaine 1987, \S 2.6), and the fraction of bulge light is generally
$\la 10\%$.  Thus the rotation curve shape from the luminous matter
changes little as the division of light between the disk and bulge is
adjusted.

Our decomposition method is described in \citet{bs03}.  Briefly, we
assume the disk to be infinitesimally thin and axisymmetric, while the
bulge is spheroidal, implying that our model isophotes for both
components should be elliptical but have differing ellipticities,
$\epsilon$.  We assume the equatorial plane of the bulge is projected
at the same inclination $i$ and position angle $\phi_0$ as the disk
plane, which are the values derived from modeling the velocity map
(\S\ref{data}).  We allow the bulge to be either oblate or prolate.
This choice varies from galaxy to galaxy, but most of our models are
oblate.  While we do not assume any functional form for the surface
brightness profile of the disk -- it is represented as a table of
values -- we follow PW and model the bulge as a sum (up to 3) of
Gaussian light distributions.  Detailed photometry of bulges with high
spatial resolution \citep[\eg,][]{bal03} suggests that Gaussians may
not be the most appropriate, but we find that a sum of three Gaussians
can reproduce, \eg, S\'{ersic} profiles \citep{ser68} reasonably well.
Though unorthodox, Gaussians have two advantages: the surface
brightness is guaranteed to decrease faster than any likely disk
profile at large radii, and they afford a convenient calculation of
the central attraction (see \S\ref{bvcalc}).  Our assumptions are less
restrictive than in schemes that utilize specific functional forms for
the luminous components.

We first attempt to fit the photometric image with a disk alone.  We
then repeat the fit with the addition of a single Gaussian bulge
component and compare the resulting $\chi^2$ value with its value from
the disk-only fit.  (The dominant source of uncertainty is simple
photon counting.)  If the difference is significant, as indicated by
an $F$-test, we continue to add a second and possibly a third
component, accepting each only if the reduction in $\chi^2$ is large
enough that there is a better than 95\% chance that the extra
parameters are required.  Our fitted photometric model
typically has final reduced $\chi_r^2$ (see \S~\ref{vffit}) in the
range $1 \la \chi_r^2 \la 1.5$.

Our best-fit model has the following parameters; a disk ellipticity
$\epsilon_D$ that is related to the inclination $i$ of the galaxy
through $\cos{i}=1-\epsilon_D$, a set of disk intensities \{$I_D$\} at
equally-spaced radii, the bulge ellipticity $\epsilon_B$, and the
Gaussian central intensities \{$I_B$\} and scale lengths \{$r_B$\} of
the bulge component(s).

\subsection{Stellar Velocity Field Calculation}\label{bvcalc}

As mentioned earlier, a Gaussian form for the bulge simplifies the
calculation of the central attraction.  Since an axisymmetric Gaussian
volume light distribution projects into a Gaussian surface brightness
distribution \citep{stark77}, the bulge luminosity density can be
derived from the sum of the bulge components through
\begin{equation}
j(m^2)=\sqrt{\frac{f}{\pi}} \sum_n
\frac{I_{B,n}}{r_{B,i}}e^{-m^2/r_{B,n}^2},
\end{equation}
where $m=R^2+(z/(1-\epsilon_B))^2$, and $f = \cos^2i + \sin^2i/q_B^2$.  
The apparent bulge axis ratio, $q_{m,B}$, is related to the bulge 
ellipticity, $\epsilon_B = 1 - q_{m,B}$ and the intrinsic axis ratio of 
the bulge $q_B$ is given by $q_B^2 = (q_{m,B}^2 - \cos^2i)/\sin^2i$. 
Finally, the bulge eccentricity $e_B$ is given by,
\[ e^2_B= \left\{ \begin{array}
  {r@{\quad,\quad}l}
  1-q^2_B & {\rm oblate} \\ 1-1/q^2_B & {\rm prolate.}
  \end{array} \right. \]
The circular velocity in the equatorial plane due to the bulge is
therefore \citep{bt87},
\begin{equation}\label{bulgev}
v^2_{c,B}(R)=4\pi G \sqrt{1-e_B^2} \mlb \int^R_0 \frac{j(m^2)m^2
dm} {\sqrt{R^2-m^2e_B^2}},
\end{equation}
where \mlb\ is the mass-to-light ratio.

We evaluate disk circular velocities using the method based on Hankel
transforms for a thin disk \citep{bt87},
\begin{equation}\label{diskv}
v^2_{c,D}(R)=2\pi \cos{i} G \mld R \int_0^{\infty}
\int_0^{\infty} \Sigma(R') J_0(kR')J_1(kR) R'k \; dk dR',
\end{equation}
where $\Sigma$ is the surface mass density, \mld\ is the mass-to-light
ratio, and $J_0$ and $J_1$ are cylindrical Bessel functions.  The
$\cos{i}$ factor is an inclination correction for the surface
brightness, that assumes a thin disk with no radial dependence to the
extinction law.  We solve Eq. (\ref{diskv}) numerically, using a
kernel approach.  Since we neglect an (unknown) thickness correction,
which would slightly reduce the circular speed for the same surface
density, our \mld\ values will be systematically underestimated.

\subsection{Velocity Field Fitting}\label{vffit}

We fit our mass model to the galaxy by comparing the projected
velocity map of our axisymmetric model to the observed 2-D velocity
map from the FP data.  As mentioned earlier, we adopt the kinematic
projection angles from \citet{bs03}, which we also used for the 
photometric
model.  We therefore have to minimize $\chi^2$ for the remaining free
parameters, which are the mass-to-light ratios for the disk \mld\,
bulge \mlb\ (if any), and parameters of the adopted halo.

The line-of-sight velocity at the $k$th-point in the projected disk 
plane predicted by our model is
\begin{equation}
v_{p,k} = v_s + v_{c,T} \frac{\sin i \cos i \cos(\phi-\phi_0)}
{\sqrt{1-\sin^2 i \cos^2 (\phi-\phi_0)}},
\end{equation} 
where $v_{c,T}$ is the total circular speed from all mass components
combined at that radius, and $\phi_0$ and $v_s$ are, respectively, the
previously-determined major axis position angle and systemic velocity.

We evaluate the quantity $z_k \equiv (v_{d,k}-v_{p,k})/\sigma_k$ at
every pixel $k$, where $v_{d,k}$ is the estimated line-of-sight
velocity derived previously by PW from the FP data.  The quantity
$\sigma_k$ is the statistical uncertainty, estimated by PW from the
Voigt profile fit to the FP data cube, plus a small additional
constant term of 7 km/s added in quadrature to represent the likely
velocity dispersion of gas in a disk.\footnote{The intensity profile
for any one pixel is broadened by the instrumental resolution, the
intrinsic line-width from an individual HII region, and by the
different velocities of the various HII regions along the line of
sight.  The Voigt fit corrects for the instrumental broadening, which
has a FWHM $\sim 125\;$km/s.  An estimated velocity uncertainty $\la 7\;$km/s
is therefore unreasonably low, and would give that datum too high a statistical
weight in the fit.  Furthermore, a single bright HII region that
dominates the spectrum in one pixel may not be as close to the
circular speed as indicated by the fitted error if it were in the
wings of the turbulent velocity spread of the ISM.}  The usual reduced
\chisq\ is therefore
\begin{equation}
\chi^2_r = \frac{1}{\nu}\sum_{k=1}^N z_k^2,
\end{equation}
where $N$ is the number of pixels in the velocity map where we have an
estimate of the line-of-sight velocity $v_{d,k}$, and the number of
degrees pf freedom, $\nu = N -$ the number of parameters.  We omit only
those pixels where the Voigt-fitter failed to find a statistically
significant velocity estimate from the FP velocity scan.

The value of \chisq\ after minimization was unacceptably large for
many galaxies.  Inspection of the residual maps in these cases
indicated that a small fraction of pixels, generally in the outer
parts of the model, had differences of many $\sigma_k$ between the
model and data.  The estimated velocities in these pixels generally
differed significantly from those in the surrounding pixels, which is
unlikely to be correct on physical grounds, and such outlying points
occasionally exerted undue influence on the fitted parameter values.
The spurious velocities in these pixels generally arise from falsely
identified ``emission lines'' that formally have small errors, which
could be caused by incompletely removed cosmic rays, for example.
Non-circular streaming motions, particularly in bars, are further
sources of major disagreement between our circular flow model
predictions and the data.

We therefore adopt a robust estimation procedure using Tukey's
biweight, which reduces the influence of non-normally distributed
errors on the fitted parameters \citep{press92}.  In this case, we
minimize the function
\begin{equation}\label{defbiwt}
\biwt = \frac{1}{\nu}\sum_{k=1}^N
\cases{z_k^2-\frac{z_k^4}{c^2}+\frac{z_k^6}{3c^4}, & $|z_k| < c$ \cr
\frac{c^2}{3}, & otherwise \cr}
\end{equation}
where $c$ is a constant.  While data values that differ from the model
prediction by $\la c\sigma_k/2$ contribute to \biwt\ almost exactly as
for the conventional \chisq, the contribution from data values that
are farther from the model prediction by $c\sigma_k$ does not change
as the parameters are adjusted, and outlying data therefore do not
influence the resulting parameter values.  \citet{press92} recommend
$c=6$, but non-circular flow patterns in our galaxies suggest that a
larger value would be more appropriate; after some experimentation we
found $c=10$ still eliminated the influence of the extreme outlying
velocities while allowing the great majority of the pixels to
contribute to the fit with almost full weight.

Since all values of $z_k$ are adjusted during the minimization, the
impact of the biweight changes at every iteration.  It is therefore
inadequate to first derive a 1-D rotation curve to which we can fit
our models; we must always fit to the entire 2-D velocity field.

\subsection{Steepness of the Inner Rise}

Many authors limit the upper value of $\ml$ so that the model
rotation curve rises no more steeply than the observed curve in the
inner parts.  We, on the other hand, try to minimize the difference
between the model and all the kinematic data without this constraint.
As a result, our model rotation curve may exceed that observed in the
central region.

We do not regard {\it mild\/} overfitting in this region as a serious
shortcoming of our models.  The reliability of the innermost data
points has been discussed extensively in the context of halo
constraints from LSB galaxies \citep{vdb00,dbmr01}, and it is argued
that every uncertainty leads to the observed velocity being lower than
the true circular velocity.  Non-circular motion, pressure support,
seeing, and patchy emission can all lead to mild underestimation of
the circular velocity in the observational data from the inner galaxy.
(We have no potential error from slit misplacement, because the
position of the rotation center is evident from our 2-D velocity
maps.)  Tests to model worse seeing gave results practically
indistinguishable from those obtained with the original seeing.
Patchy emission that is blurred by the seeing causes faint emission
with the wrong velocity at points where no signal should have been
detected.

Of these further sources of uncertainty, non-circular motions appear to
have the greatest impact.  Most galaxies in our sample have spiral
patterns and some are clearly quite strongly barred.  We have not, in
this work, attempted to model the non-circular motions induced by such
non-axisymmetric features, and continue to search for the best-fitting
model with a circular flow pattern.  This simplifying assumption
clearly boosts \biwt, but it can also introduce systematic errors.  In
particular, streaming motions in a bar generally cause the observed
orbital speeds to be lower than circular, because the gas spends more
time, and is more easily detected, near the apocenter of the streaming
pattern where the speed must necessarily be lower than circular.

For these reasons, we consider a mild overfit in the central regions
an acceptable result, and we accept a more significant overfit in
galaxies having bars or other strong non-axisymmetric features.

\section{Stars-only Fits}\label{bfits}

We first fit the velocity maps with models based on the light
distribution alone (stars-only fits).  The fitting parameter is \mld,
with a second \mlb\ parameter when a bulge is included.  We list the
best-fit parameters along with their uncertainties (estimated as
described below) in Table \ref{btab}.

\begin{figure}
\plotone{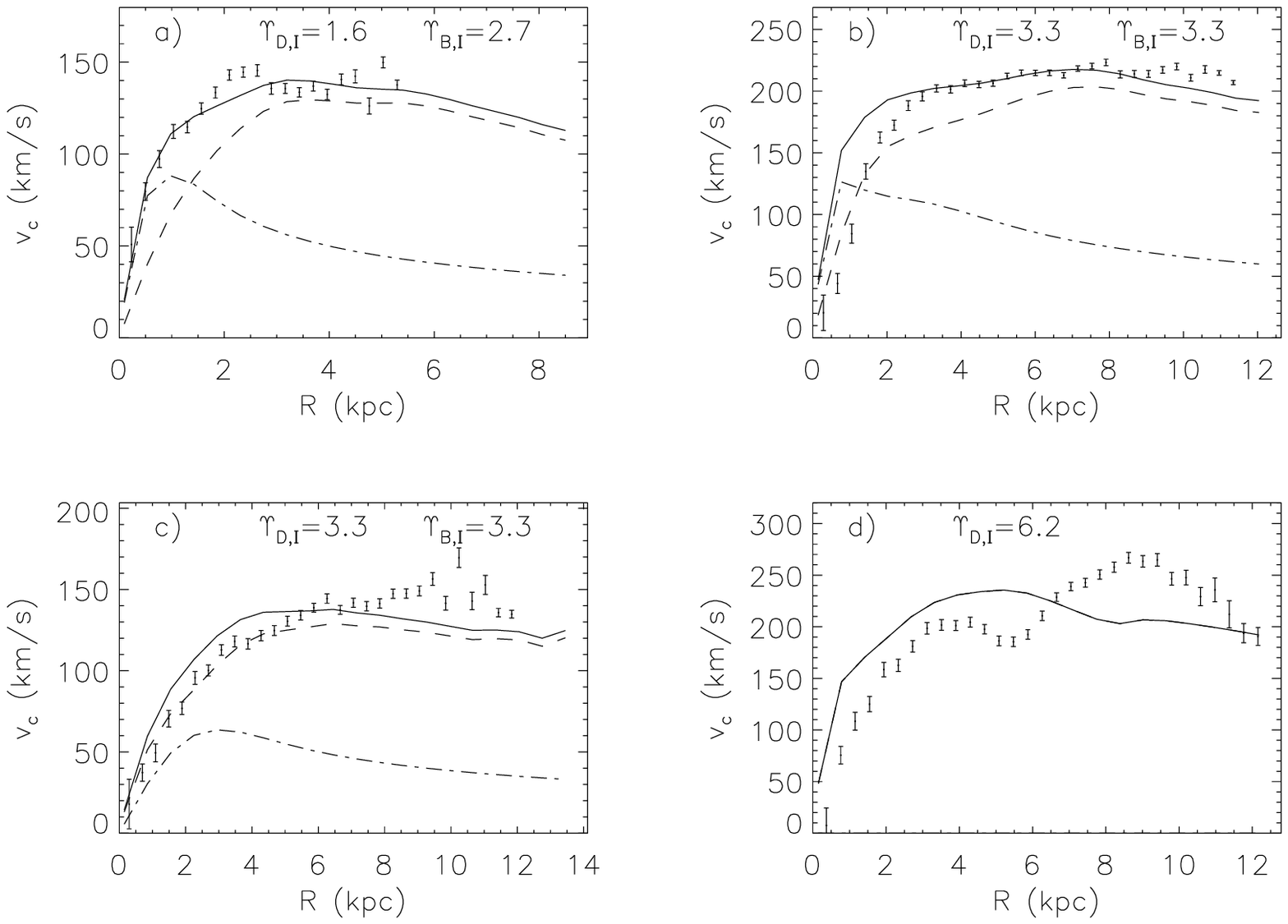}
\caption{Various types of
rotation curves for the stars-only models.  The dashed line represents
the disk contribution.  The bulge contribution is shown with the
dash-dotted line.  The solid line is the combination disk+bulge. a)
ESO 268g44: an example of good agreement between model and data.  b)
ESO 439g20: a barred galaxy for which the model overpredicts the inner
rotation curve; the bar has a half-length of approximately 2 kpc.  
c) ESO 444g47: an example of an overfitted disk
driven by an evident mass discrepancy in the outer regions.  d) ESO
381g05: a galaxy with strong spirals and a possible tidal distortion.
\label{borcs}}
\end{figure}

Figure \ref{borcs} shows our results for four representative galaxies.
The mean orbital speed as a function of radius is shown by the points
with error bars; these values are deduced from the 2-D FP velocity map
by the method described in \citet{bs03}.  The solid line in these
Figures shows the rotation curve of our best-fit, stars-only, model;
the disk contribution is the dashed line, and the bulge contribution
(if present) is the dash-dotted line.  The four panels in Figure
\ref{borcs} show examples of different types of behavior.

PW found that the large majority of galaxies in their sample could be
modeled adequately without dark matter.  Not surprisingly, we have
also found that reasonable \mlb\ and \mld\ values can describe the
data well in 34 of our 40 galaxies; Figure \ref{borcs}(a) shows a good
example (ESO 268g44).  In some cases (\eg, ESO 438g15 and ESO 502g02)
the model predictions begin to drop below the data in the outer few
points, but match the shape of the inner data.

The fit in the inner parts of eleven out of these 34 galaxies is
significantly worse, due to bars and/or strong spirals that invalidate
our assumption of circular orbits, but the amplitude and shape of the
rotation curve over some radial range outside the barred region is predicted quite well by
our mass model.  Figure \ref{borcs}(b) shows an example of this
behavior (ESO 439g20) -- the model rotation curve runs significantly
above the data values in the barred region, which is typical of these
11 cases.

Five of the remaining six cases exhibit the behavior seen for ESO
444g47 (Figure \ref{borcs}c), in which there is no significant radial
range where the prediction matches the data.  The predicted rotation
curve shape begins to drop well before the optical edge, whereas the
data do not; this shape difference is a clear indication of a mass
discrepancy within the visible disk.  With no dark matter component
available, the fitting routine returns an unreasonably large \mli\ in
order that the underprediction farther out is balanced by a
compensating overprediction of the observed speed in the inner parts.
Since the \mli\ is larger than can be allowed by the inner rotation
curve in these five cases, we follow PW
and eliminate the outer data to find the maximum allowable \mli.
Subsidiary fits in these five cases are also reported at the end
of Table 2.

The remaining galaxy, ESO 381g05, shown in Figure \ref{borcs}(d), has
large variations that are not well-predicted by the photometric model.
This galaxy has a single strong spiral on one side, and may be
disturbed by a recent, or on-going, merger.

The total stellar masses we derive agree quite well with those found
by PW from the same data.  The agreement is not perfect for several
reasons: In all but five cases, our fits include all the kinematic data
whereas PW limited their fits to the inner parts for all galaxies with
outer mass discrepancies.  Our disk-bulge decompositions are not
identical and we also required $\mlb > 0.5\mld$ whereas PW frequently
assigned much less mass to the bulge light.  But the largest
differences arise because we use kinematically defined inclination
angles, whereas they preferred the inclination derived from the
photometric image; in a few cases, this difference exceeds $10^\circ$,
leading to quite substantial differences in \mli.  In fact, the
largest discrepancy is for ESO 381g05, which may be tidally disturbed,
calling the assumption of an intrinsically flat disk into question.

\subsection{Uncertainties}

The uncertainty in the fitted \mli\ values is due mostly to the
uncertainties in the galaxy's inclination, position angle, and
distance.  (Since we fit our models to a large number of data points,
the \mli\ values are tightly constrained once these parameters are
fixed.)  In a previous paper \citep{bs03}, we estimated systematic
errors in inclination and position angles caused by spirals and other
nonaxisymmetric structure in the galaxies.  The best-fit value plus
the estimated high and low limits for each of the two projection
angles yield nine different combinations of inclination and position
angle which give nine separate estimates for \mli\ values.  We take the
uncertainty in an \mli\ value to be half the range of these nine \mli\
values.

Since \ml\ is inversely proportional to distance, errors in our Hubble
distances due to peculiar motions will also bias our estimated values.
Indeed, this sample of galaxies was originally observed to study
peculiar velocities in the region toward the ``Great Attractor'' and
may therefore have larger than usual peculiar velocities.  Since
typical redshifts are 4000--5000 km/s, relative distance errors are
unlikely to exceed 25\%, however.  \citet{bothun92} use the
Tully-Fisher relation to estimate distances, finding differences
between redshift distance and TF distance of typically about 25\%.
They find that the majority of these galaxies are actually closer than
their Hubble distances, implying that our \mli\ values may be
underestimated by this factor.

\subsection{Discussion}

For our sample, the average \mld\ is 2.6 and the average \mlb\ is 2.8
(using the \mli\ values of models with the best-fit $i$ and $\phi_0$).
The average \mld\ and \mlb\ systematic uncertainties are
$s_{\mld}=0.5$ and $s_{\mlb}=0.7$.

Simulations by \citet{bdj01} predict that the mass-to-light ratio of a 
stellar population should correlate with 
its intrinsic $B-R$ color.  \citet{bell03} have applied
the Bell \& de Jong models to a large sample of galaxies with multi-band 
photometry and find a somewhat shallower relation, with a large scatter, 
which we show by the line in Figure~\ref{barycolplot}(a).
We plot our stars-only estimates of \mld, but it should be noted
that these dynamical estimates are only upper limits -- the stellar
\mld\ could be lower if dark matter contributes significantly to the
central attraction in the inner galaxy.  The error bars on
\mld\ are the systematic uncertainties discussed above, and the arrows
indicate how \mld\ would change if the TF distance were used in place
of the Hubble distance.

As we have only $I$-band images, we use galaxy colors obtained from
the ESO-LV aperture magnitudes, as recorded in NED.  In principle, we
should compare the \mld\ with the color of the disk alone, but NED
gives magnitudes of the disk and bulge combined.  The bulge light in
our mostly late-type galaxies is a small fraction, typically $\leq
10\%$, of the total in the $I$-band, so any correction for the bulge
to the total color is likely to be small.  We have corrected the NED
magnitudes for Galactic extinction using values from \citet{schleg98}
and for internal extinction using the prescriptions of
\citet{tully98}.  Errors in our estimated colors might be $\sim
0.15\;$mag.

Our average \mld\ values (and those of PW) are in reasonable agreement
with those found by \citet{bell03} and it is interesting that there are no
points significantly below their trend.  A point lying above the
line, on the other hand, could simply indicate a
significantly submaximal disk.  While this could be the reason we see
no trend, we first explore whether other effects could mask a trend in
stars-only \mld\ with color.

It is extremely unlikely that the corrections for the bulge light to
the total colors could be large enough to destroy a trend, and the
galaxies with no bulge (square symbols) also do not show one.
Unfortunately, the number of galaxies without bulges is too small for
a 2-D Kolmogorov-Smirnov test to determine whether this group is
distributed differently from the bulge-and-disk galaxies in
(\mli,color) space.

It is possible that a trend is masked by large errors in our \mld\
values.  The error bars include our estimate of the systematic error
arising from uncertainties in the projection angles but do not include
possible distance errors.  They would have to be very large to mask
the trend entirely, or conspire to correlate with color, which seems
unlikely.

\citet{bell03} find the \ml\ values of blue stellar populations are 
more sensitive 
than red to metallicity.  It is possible that the bluer galaxies in our 
sample are all metal-rich, but metallicity correlates with luminosity
and we see no obvious trend of luminosity with color.

We are suprised not to see any trend of stars-only \mld\ with color,
since a trend has been seen in other work \citep[\eg,][]{mcg04}.  The
absence of a trend may imply that the halo contribution must correlate
with color, which seems contrived for these bright HSB galaxies.  The
points in Figure \ref{barycolplot}(b) are encoded in many ways to
distinguish galaxies by luminosity, field galaxies from those in
clusters, the membership of different clusters, and the arrows show
where the point would move if we adopted photometric inclinations,
instead of kinematic.  None of these appears to be a factor that could
account for the absence of a trend.

\section{Dark Matter Halo Fits}\label{dmfits}

We next report fits to the velocity map that include a dark matter
halo component.  Since our adopted halo models have two free
parameters, we now have to minimize \biwt\ for either three or four
parameters, depending on whether a separate \mlb\ is used.

Estimating statistical uncertainties from the $\biwt$ surface in a
4-parameter hypervolume is technically more difficult than for the
2-parameter stars-only models.  The difficulty is compounded because
we often find that the adopted halo parameters are strongly degenerate
-- that is, the minimum in the \biwt\ hypersurface lies in a long,
curved, and almost flat-bottomed valley.  We estimate the
uncertainties in each fitted parameter using a Markov Chain Monte
Carlo method \citep[\eg,][]{chris01,kmj02}, deriving the estimated
likelihood of a set of parameters from the \biwt\ value.  Our Markov
Chains for each galaxy contain 1~million elements and appear to be
well-converged.  We construct a histogram of the chain elements in the
multi-dimensional parameter space; the likelihood region is the set of
all bins with more than some threshold number of chain elements, with
the threshold chosen such that the selected bins contain 68\% of the
chain elements.  The set of chain elements in the selected likelihood
region can then be projected onto the parameter axes to determine the
statistical uncertainties.  Figure~\ref{mchist} shows an example for
the Pseudo-isothermal (PI) halo model of ESO 322g82.  Strictly
speaking, these confidence regions are not exactly the 1-$\sigma$
uncertainties, since we define likelihoods from \biwt, and not the
usual \chisq\ function.  However, we find in practice that working
with \chisq, in cases where the outlying points have little effect on
the position of the minimum, leads to very similar confidence regions.

We use an $F$-test to determine whether a halo is required to fit our
data.  This test determines whether the reduction in \biwtm\ from the
stars only case to when the fit includes the extra two halo parameters
is large enough to justify the additional parameters.  (Again, we
treat \biwt\ as an adequate surrogate for \chisq\ and, of course, we
always compare with \chisq\ for the entire velocity map from the
stars-only fits.)  We discard the fits in a few cases only where this
test indicates that there is less than 95\% chance that the halo is
necessary, since the values of the parameters in such cases are
unlikely to be meaningful.  Just three galaxies (ESO 317g41, ESO
382g06 \& ESO 438g08) fail this test for all the types of halo we have
fitted.

We also find in many cases that the halo drives \mld\ to very
low values, and the observed flow pattern is determined largely by the
halo.  It is clear that $\mld \ll 1$ is unphysical -- the stellar
population of the disk clearly must have some significant mass.  Since
the galaxies in our sample have quite normal colors, we regard $\mld <
0.5$ to be unrealistically low, and therefore discard any fits for
which the 1-$\sigma$ upper bound on $\mld < 0.5$.  Very few
cases fall just below this cut, and $\mld \ll 0.1$ for the large
majority discarded for this reason.  This criterion leads us to
discard one galaxy in our sample (ESO 323g39) for all types of halo we
have tried.

We have not attached any significance to the actual \biwtm\ value,
which always exceeds unity and is mostly greater than 2 -- ranging up
to greater than 11 in one case!  Formally at least, large values of
this parameter indicate that the model is an unacceptable fit to the
data, which we do not dispute.  However, examination of the residuals
in the 2-D velocity map indicates that the large \biwtm\ values arise
from three principal sources: outlying points, bars, and spirals, in
decreasing order of importance.  We argue that large values of \biwtm\
do not invalidate our estimates of halo parameters.  The biweight
function is a robust statistic designed to minimize the influence of
the outlying points, even though they still make a large contribution
to the value of \biwtm\ (eq.~\ref{defbiwt}).  The halo parameters are
determined more by the overall shape of the rotation curve than by the
non-axisymmetric streaming motions, which could possibly be modeled
with a lot of effort (\eg\ Weiner et al.\ 2001b), but do not affect
the final parameter values to a great extent \citep{wsw01a}.

\subsection{Pseudo-isothermal Halos}

The PI density distribution is
\begin{equation}\label{isorho}
\rho_{\rm PI} = \frac{\rho_0}{1+(r/r_c)^2},
\end{equation}
specified by two parameters $\rho_0$, the central density, and $r_c$,
the core radius.  The associated squared circular velocity is
\begin{equation}\label{isovel}
v_{c,\rm PI}^2 = v_\infty^2 \left[1-\frac{r}{r_c}\arctan\left(
\frac{r}{r_c}\right) \right],
\end{equation}
where $v_\infty \equiv (4\pi G \rho_0)^{1/2} r_c$ is the asymptotic
circular velocity.

The best-fit parameters and their associated uncertainties for all 40
galaxies are given in Table \ref{itab}.  In only 3 cases out of the 40
galaxies in our sample does the addition of the extra two parameters
for the halo lead to such a small reduction in \biwtm\ that the halo
is not needed; the value of \biwtm\ in these cases is marked with an
asterisk; we discard these three cases in the following discussion.

Ten of the remaining 37 galaxies have \mld\ values for which the
1-$\sigma$ range never exceeds 0.5. Non-circular motions in the bar
may be one of the factors that drives the solution towards a low \mli\
value and a dominant halo.  In these cases, the circular speed is
often underestimated, while the rotation curve predicted by the
luminous matter has a more steeply rising shape.  Thus the fit can be
improved by decreasing the \mli\ and fitting the data mostly with the
halo, which has a more adjustable shape.  Whether this happens depends
on other factors also, such as the bar orientation.

The halo parameters ($\rho_0$ and $r_c$) of the remaining 27 galaxies
span a wide range, as shown in Figure~\ref{pivals}.  The line of slope
$-2$ in this plot is not a fit to the data, but shows the relation
$v_\infty^2=4\pi G \rho_0 r_c^2$ for $v_\infty = 200\;$km/s.  Galaxies
falling above (below) this line have larger (smaller) values of
$v_\infty$.  The fitted points are not randomly distributed about the
line; the points above (below) the line tend to have large (small)
$r_c$ values.  The distribution suggests, perhaps only marginally,
that larger $v_\infty$ galaxies, which also tend to be more luminous,
have larger, lower-density cores than do the smaller galaxies.
Despite using a slightly different cored dark matter distribution, the
6 galaxies presented in \citet{gen04} follow essentially the same
relation between central density and core radius.  \citet{don04} also
present isothermal fits for 25 galaxies.  Ten are ``regular'' spirals
(neither dwarf nor low surface brightness), and they too lie among our
points in the ($\rho_0$,$r_c$) plane.  \citet{don04} have additionally
found a tight correlation between the disk scale length and $r_c$.  PW
fit exponential disks to their galaxies and determined disk scale
lengths ($R_d$), and we adopt these values as reasonable estimates.
Despite the trend for the galaxies with larger $v_{\infty}$, and
presumably larger disks, to have larger $r_c$ values, we do not
reproduce their claimed correlation between $R_d$ and $r_c$.
\citet{kor03} find $\rho_0 \propto r_c^{-1.2}$ for PI halo fits, which
is a shallower slope than the trend shown our Figure~\ref{pivals}.
However, their fits are made with the assumption of a maximum disk,
which precludes central densities as high as some that emerge from our
fits.

The three error bars that extend from points well beyond the
right-hand edge are from three cases (ESO 322g45, ESO 323g73 \& ESO
375g02) in which halo core radius is huge and very poorly constrained;
the halo contribution to the circular speed rises essentially linearly
to the outermost observed velocity.  We are therefore unable to
determine $v_\infty$ with any confidence.  Such degeneracy has been
reported by other authors, even by \citet{bro92} who had high quality
HI data extending to radii well outside the optical disk.

\subsection{NFW Halos}

The collapse of dark matter halos in simulations of large-scale
structure formation yields halo density profiles that are not at all
like the pseudo-isothermal function used above.  The precise form of
the density profile that emerges in these simulations is still
disputed \citep{nfw96,power03,dms04}, but all agree that the density
increases steeply towards the center.  We have therefore tried a broken
power-law for the halo density profile.  In this work, we treat the halo as
spherical and make no attempt to take account of halo compression as a
result of the cooling and settling of baryons to form the luminous
component.  Our objective is simply to assess whether the data are more
consistent with a cusped density profile in galaxies today, and not
to test detailed predictions from cold dark matter (CDM) simulations.

We adopt the original NFW density profile:
\begin{equation}\label{nfwrho}
\rho_{\rm NFW}=\frac{\rho_s r_s^3}{r(r+r_s)^2},
\end{equation}
where $\rho_s$ is a scale density and $r_s$ is a scale length.  The
squared circular velocity arising from this density profile is
\begin{equation}\label{nfwvel}
v_{c,\rm NFW}^2=v_{200}^2 \left[ \left(\frac{c}{x}\right) 
\frac{\ln(1+x)-x/(1+x)}{\ln(1+c)-c/(1+c)}\right],
\end{equation}
where $x\equiv r/r_s$, $c\equiv r_{200}/r_s$ is the concentration
parameter and $v_{200}=10c r_s H_0$ is the circular speed at
$r_{200}$.  Inside the radius $r_{200}$ the average density is 200
times the critical density of the universe ($3H_0^2 / 8\pi G$).

The best-fit NFW parameters ($c$ and $v_{200}$) are listed in Table
\ref{ntab}.  The reduction in \biwtm\ is too small to justify the two
halo parameters in 4 cases out of the 40 galaxies in our sample; the
value of \biwtm\ in these cases is marked with an asterisk.  Once
again, we discard these for the remainder of this section.  We also
discard 16 of the remaining 36 galaxies because the 1-$\sigma$ range
of their \mld\ values never exceeds 0.5.

As with the pseudo-isothermal halo fits, the best fit NFW halo
parameters of the remaining 20 galaxies have a considerable range, as
shown in Figure~\ref{nfwvals}.  In only 9 cases, which tend to have
higher values of $c$, the 1-$\sigma$ range of $v_{200}$ does not
extend up to values greatly in excess of any observed velocity.  The
enormous error bars for the other 11 galaxies indicate that the halo
parameters are strongly degenerate.  The line is the mean correlation
between these halo parameters in a concordance $\Lambda$CDM simulation
\citep{bull01}.  While our model halos are not directly comparable to
those from simulations (see \S~\ref{hden} for a more thorough
discussion of the comparability of the simulated and model halos) and
the uncertainties are quite large, the points in Figure~\ref{nfwvals}
appear to describe a steeper trend than the prediction.

\subsection{Power-law Halos}

The results from fitting the above two standard halo models were not
particularly informative for a number of reasons.  Extreme minimum
disks are often favored in a significant fraction of the galaxies
(mostly those with strong non-axisymmetric features), which we then
discard as unphysical.  Also, the halo parameters are frequently
highly degenerate since the minimum of the \biwt\ hypersurface lies in
a long, narrow, curved valley.

However, rotation curve data do tightly constrain something, namely
the mass or mean density interior to the last measured point.  The
just-noted degeneracy in the NFW parameters arises because halos
having the allowed wide range of parameters all have closely similar
density profiles in the inner region that is the only part
constrained by our data.  Cases where no mass is assigned to the
luminous material appear to arise because fitting functions with a
length scale and a density scale as free parameters have too much
freedom to adjust to quirks in our data.

We have therefore also tried a halo density profile with a smaller
range of possible shapes: a single power law with no characteristic
length scale.  The density profile we adopt is
\begin{equation}
\rho_{\alpha}(r)= \rho_l \left( {r_l \over r} \right)^\alpha,
\end{equation}
where $\rho_l$ is the halo density at $r_l$, the radius of our
outermost velocity measurement.  We stress that we fit this function
only to the radial range spanned by our data and cannot say how the
density profile continues beyond this radius.  The squared circular
velocity in this halo is
\begin{equation}
v_{c,\alpha}^2=v_l^2 \left(\frac{r}{r_l}\right)^{2-\alpha}.
\end{equation}
Again we have two free parameters, $v_l$ and $\alpha$, with
$\rho_l=(3-\alpha)v_l^2/(4\pi Gr_l^2)$.

Table \ref{ptab} gives the best-fit parameters and uncertainties for
the power law fits.  The reduction in \biwtm\ is too small to justify
the two halo parameters in 6 out of the 40 galaxies in our sample; the
value of \biwtm\ in these cases is marked with an asterisk.  Once
again, we discard these.  We need to discard only 2 of the remaining
34 galaxies because the 1-$\sigma$ range of their \mld\ values never
exceeds 0.5.  This small fraction contrasts with the 10 and 16 cases for
the PI and NFW fits respectively.

With this halo form, we do not find strong degeneracies between the
fitting parameters, implying that the fitted values are much more
robust.  Figure~\ref{powvals} shows that the halo parameters are also
spread over a more limited range.  Yet despite the more limited range
of possible shapes, the values of \biwtm\ are scarcely significantly
larger than for the more flexible functions, and are actually smaller
in a few cases.

The reason the power-law halo model is more useful, as judged by the
preceding statistics, is that it has less flexibility.  The length
scale in the pseudo-isothermal and NFW halo models can produce a
feature in the halo contribution that matches a feature in the data,
thereby yielding a lower \chisq, even though this is often achieved
at the expense of driving \mld\ to an absurdly low value.  In these
cases, it is likely that some non-axisymmetric disk feature drives the
fit to this part of parameter space, rather than a true feature in the
halo profile.

Figure \ref{powvals} shows that the distribution of $\alpha$ values is
concentrated between $0.8 \la \alpha \la 1.6$, although a few values
are close to zero and some range up to $\alpha \la 2$.  The
significance of $\alpha \ga 1$ is unclear, however, since the probable
halo contribution should be something in between uniform rotation and
exactly constant orbital speed, which requires $0 < \alpha < 2$.
While all intermediate values yield a halo rotation curve that rises
fast initially and bends over to a less steeply rising part towards
the outer radius, the abruptness of this transition does increase with
$\alpha$, and the concentration of values in the range $0.8 < \alpha <
1.6$ may be significant.  Recent cosmological N-body simulations have
also found similar $\alpha$ values \citep{dms04}.

\section{Disk Masses and Halo Densities}

\subsection{Mass-to-light Ratios}

We plot the \mld\ values of the three halo models versus color in
Figure \ref{halocolplot}.  The error bars now reflect only the
statistical uncertainties in \mld\ values, and do not include any
estimates of deprojection uncertainties which are of comparable
magnitude, but which would be burdensome to compute.  Since the colors are the same,
the abscissae and their uncertainties in each panel are the same as in
Figure \ref{barycolplot}.  Each panel of this figure has fewer than 40
points because we have discarded any fit for which the halo was
unnecessary or that yielded such a low \mld\ value that its 1-$\sigma$
range never exceeded 0.5.  We again indicate the relation fitted by
\citet{bell03}.  The most striking feature of these plots is that
adding halos has not changed the flat character of the distribution of
points seen in Figure~\ref{barycolplot}; we still do not see a trend
with color for any of the three halo forms.

Our velocity data in a few cases do not require a halo component at
all.  This statement is true for all halo types in just three
galaxies: ESO 317g41, ESO 382g06, and ESO 438g08.  
ESO 317g41 and ESO 438g08 have strong spirals, but ESO
382g06 is axisymmetric.  The kinematic data for ESO 317g41 extends
only to 80\% of ${\rm R}_{23.5}$, the radius at which the $I$-band
surface brightness is 23.5 mag/arcsec$^2$.  This raises the
possibility that the radial extent of our data too small to probe the
halo.  However, for ESO 382g06 and ESO 438g08, the kinematic data
extend at least to ${\rm R}_{23.5}$, approximately 4 disk scale
lengths.  The rotation curves of these two galaxies turnover and
decrease in the outer parts.  As these galaxies appear to be baryon
dominated, one might expect good agreement between our
\mld\ values and those predicted by \citet{bell03}.  In fact, we find
that only ESO 438g08 is marginally consistent with the predicted value
while the other two are several sigma above the prediction.
Even though none of the simple halo functions tried here
improves the fits by enough to justify the extra parameters, these three
galaxies may yet have dark matters halos.  In fact, if the halo
rotation curve has a similar shape to that of the disk in the range
covered by our data, any relative fraction would yield an equally
acceptable fit.

Some galaxies, such as ESO 216g20 -- the reddest in our sample, do
seem to require a halo that makes their fitted \mld\ fall below that
expected for a stellar population of its color.  We note that the
TF-distance for this galaxy is larger than its Hubble distance, so its
unusually low \mld\ is unlikely to result from too short an adopted
distance.  ESO 322g44 is another case which falls below the line
in all three panels and there are others which are low when we accept
the fit, and are discarded as unrealistically low for other halo
types.

Of the galaxies lying above the line from \citet{bell03} in Figure
\ref{barycolplot}, only one of the bluest galaxies in our sample, ESO
446g01, remains significantly above the line in all the
\mli\ vs. color plots, while three other galaxies (ESO 381g05, ESO
437g31, ESO 501g01) lie well above the line for the type(s) of halo(s)
for which we accepted the fit, but are discarded for other types.  The
introduction of any of three halo types into these four galaxies did
not reduce their \mld\ values to the predicted value.

None of our adopted halo models yields \mld\ values that follow the
trend found by \citet{bell03}.  It is possible that systematic errors
such as distance and/or inclination errors, radial extinction variations, 
gas mass corrections, thickness corrections, non-circular motions, etc., 
have masked the expected trend, but they would have to be as large as 
factors of three, and in a sense to correlate with color.  It is more 
likely that our models have mis-estimated the dark matter content in 
the majority of these galaxies, and our \mld\ values are spurious.

Figure \ref{mlcompare} compares the \mld\ values obtained from each of
our fits to the 2-D velocity maps.  Each panel displays considerable
scatter.  The values from adopting one halo model are frequently
inconsistent with those from another model at the several-$\sigma$
level, although a few cases, which generally have smaller
uncertainties, do lie nicely on the diagonals in all three panels.
The open circles mark galaxies which fail the $F$-test for either of
the halo models.
The most significant feature in this Figure is that the NFW halo
density profile generally leads to the lowest assigned \mld\ value of
the three functions we have tried.

\subsection{Halo Masses}

We have determined the total mass in the halo interior to our
outermost available velocity measurement.  The halo masses for the
three different functional forms, normalized by total enclosed mass,
are compared in Figure~\ref{maptomap}.  Again, the open circles mark
galaxies which fail the $F$-test for either of the halo models.  We
have used our MCMC study to determine the confidence intervals, which
range up to 70\% in the worst few cases but are generally much lower.

It is clear from this Figure that the NFW form tends to require a
larger halo mass fraction than the other two types, which is
consistent with its generally lower disk masses.  Many cases require
no halo or negligible disk mass, and other cases show large
differences in fitted halo masses between the different halo models.

In general, the agreement between the halo masses calculated for the
different halo types is poor.  For 12 galaxies, the total uncertainty
in PI halo mass fraction is $<0.2$.  These galaxies are, on average,
$2.3\sigma$ away from agreeing with the NFW halo mass fractions and
are $2.1\sigma$ away from agreeing with the power-law halo mass
fractions.  The 5 galaxies which have NFW halo mass fraction
uncertainties $<0.2$ are $1.3\sigma$ and $2.1\sigma$ from agreeing
with the PI and power-law values, respectively.  Finally, 17 galaxies
have power-law halo mass fraction uncertainties $<0.2$.  These
galaxies are $2.2\sigma$ and $2.3\sigma$ away from agreeing with the
PI and NFW values, respectively.  However, in a few cases the halo
mass fraction has a small uncertainty, is neither 0 nor 1, and lies
close to the diagonal in this Fig.~\ref{maptomap}.  One such case is
ESO 445g15, for which we claim a consistent and well-constrained halo
mass fraction for all three density profiles we tried.

Aside from these rare cases, we conclude that not even our 2-D data
constrain the dark matter content of a galaxy -- the fitted amount is
either highly uncertain, or depends on the halo mass function adopted.

\subsection{Halo Densities}\label{hden}

Cosmological simulations make predictions for the central densities of
dark matter halos.  Specifically, \citet{alam02} introduce a measure
of the central densities of halos, $\Delta_{v/2}$, which is the
average halo density inside the radius at which the velocity is half
the maximum halo velocity [$\rho(R_{v_h/2})$], normalized by the
critical density of the universe, $\rho_c=3H^2_0/(8\pi G)$.  Their
Figure 2 plots the predicted $\Delta_{v/2}$ from an $\Lambda$CDM
simulation versus maximum halo velocity $v_{\rm max}$ in which larger
dark matter halos (high $v_{\rm max}$) have lower densities.  The same
figure has points derived from real galaxies which do not appear to
follow the predicted trend.

We show our derived $\Delta_{v/2}$ versus $v_{\rm max}$ for the NFW
halos only in Figure \ref{simcomp}.  The line illustrates the trend
expected from simulations \citep{zent02}.  One point marked by an open
circle is one of the four cases for which the $F$-test indicates no
halo is required, two other cases are points off the left side of the
plot, and the fourth case cannot be plotted as the best fit assigns
zero halo mass.  The points without symbols, which lie mostly above
the line, are fits with very low \mld.  The rest are marked with
filled symbols.  Those that have enormous error bars are cases that
are degenerate, because our optical data do not extend far enough to
constrain $r_s$.

Our data points are not directly comparable with predictions of 
$\Lambda$CDM, since the simulated halos should be compressed by baryonic 
infall.  To make a detailed test, we should include halo compression as 
a step in the minimization loop in order to determine the parameters of 
the uncompressed halos, possibly also testing a range of uncompressed 
density profiles as part of the uncertainty in the predictions.  Our 
less ambitious approach is simply to test whether the NFW profile 
describes galaxy halos in the local universe any more closely than the 
others.  Thus our parameters, if they have any relevance at all, are for 
compressed halos.  It is clear that the mean densities of the 
uncompressed halos would be lower, by an amount that depends on the 
stellar mass.

The large error bars on many points make it difficult to conclude much
from this Figure; we cannot claim that our data are inconsistent with
the prediction, but neither do they offer much support for it.  This
result is similar to the outcomes in other studies that compare
simulations to parameters derived from galaxies \citep{alam02,zent02}.
Even points with small error bars have a large scatter about the line,
whereas the simulations predict a rms scatter of about a factor 3
\citep{alam02}.  We also note that as points without symbols represent
fits that attribute essentially no mass to the luminous matter, more
realistic fits would attribute a lower average halo density and move
those points to lower $\Delta_{v/2}$ values.

We cannot make a similar plot for the power law halo, since we know 
nothing about how the profile may vary at radii beyond our data.  
However, we did obtain the lowest halo mass with this function 
(Fig.~\ref{maptomap}) which, combined with the greater compression from 
the more massive disks suggests that the densities are considerably 
lower than for the NFW halos.

\section{Rotation Curve Fitting}\label{longslit}

In this section we compare our results from the 2-D
velocity maps with what could be achieved for the same galaxies from
1-D longslit data.  As we do not have separate longslit data, we
extract pseudo-data from the 2-D velocity map along a line across
the major axis of the galaxy.

\subsection{Pseudo-slit Rotation Curves}

We mimic data that could result from longslit spectroscopy, by placing
a ``slit'' along the photometric major axis of each galaxy and passing
through the point of brightest continuum emission.  We extract the
mean velocity, $v_{\rm los}$, and its uncertainty from our 2-D
velocity maps every 1.5\arcsec, a typical seeing length, from the
pixel whose center is closest to the chosen line; interpolation or
averaging over several pixels is unnecessary, since the pixel sizes
are generally less than the seeing.  The S/N in any one pixel of our
map is probably less than would be obtained from a longslit
observation, but not by a large factor.  (The FP scan used 10 min
exposures at each filter setting, but as the device used has about twice the
throughput of a typical spectrograph, our S/N may be equivalent to
that from a 20 min spectrographic observation.)

We treat these data by the typical procedures for longslit data
\citep[\eg,][]{dbmr01}.  We estimate the systemic velocity by folding
the $v_{\rm los}$ curve, by eye, assuming that the receding and
approaching sides are symmetric.

Since a velocity map allows us to determine $v_s$ with greater
precision, we have investigated the differences between the $v_{s,\rm
map}$ and $v_{s,\rm slit}$ values.  The rms difference between the
systemic velocities derived from the pseudo-slit data and our velocity
maps is $\Delta_{\rm sys} = 10$ km/s.  It is interesting to break this
comparison down for axisymmetric, spiral, and barred galaxies.
Axisymmetric galaxies (disks without strong spiral structure) have
$\Delta_{\rm sys}=6$ km/s.  Galaxies with spirals have $\Delta_{\rm
sys}=13$ km/s.  Overall, barred galaxies have $\Delta_{\rm sys}=10$
km/s.  We further subdivide the barred galaxies according to the
position angle of the bar.  When the bar lies along the minor axis,
$\Delta_{\rm sys}=8$ km/s.  Galaxies with bars at an oblique angle to
the major axis have $\Delta_{\rm sys}=12$ km/s and those with major
axis bars have $\Delta_{\rm sys}=10$ km/s.  In the central regions of
these galaxies where typically the rotation curves rise approximately
linearly, the uncertainty in $v_{s,\rm slit}$ relates directly to an
uncertainty in the position adopted for the center.  Using the maximum
uncertainty quoted above leads to an uncertainty in dynamical center
positions $\la 1\arcsec$.

We retain all the points (receding and approaching) as independent
data points in the rotation curves.  Finally, we correct the rotation
speeds for inclination using the photometric value since this is all
slit observers would have at their disposal.

\subsection{Comparison of Map and Pseudo-slit Rotation Curves}

Our pseudo-slit rotation curves generally agree well with those we
derived from the 2-D velocity maps.  A few illustrative cases are
shown in Figure \ref{rccomp}; the dotted error bars denote pseudo-slit
data, while the solid points are from our fits to the 2-D maps.  Panel
(a) shows an example (ESO 322g82) of good agreement between the two
curves.  The case in panel (b) shows a galaxy with strong spirals (ESO
322g36); velocities differ between the approaching and receding sides
over the radial range from 3 to 8 kpc, but the velocity map rotation
curve nicely bisects this asymmetry.  The agreement between the two
curves in panel (c) (ESO 438g08) is poor, which results from a
disagreement between the photometric major axis that a slit observer
would use and the major axis that we find from our 2-D velocity map;
the spirals present in this galaxy have biased the estimate of
photometric position angle (see \citet{bs03} for a fuller discussion
of this effect).  The case shown in panel (d), ESO 501g01, shows the
greatest advantage of velocity map data over slit spectroscopy, since
this galaxy has major axis H$\alpha$ emission over only a limited
radial range, whereas the velocity map allows a much greater radial
range to be explored.

\subsection{Mass Models}

We have used our pseudo-slit rotation curves to determine \mli\ values
in a manner similar to the 2-D data.  We fit constant \mli\ models,
to the inclination-corrected velocities, using the same procedure as
described above, but for this much smaller number of pixels.  As
before, we first fit models without halos and then include the three
different halo types.

\subsubsection{Stars-only Fits}

The comparison of pseudo-slit and map \mli\ values is illustrated in
Figure \ref{rcmlcomp}.  The pseudo-slit and map \mld\ values have rms
fractional differences of 22\% while the \mlb\ values differ by 39\%.
We find that in 28 of 40 galaxies the pseudo-slit \mld\ values are 25\%
lower than the map \mld\ values.  The remaining 12 galaxies have map
\mld\ values which are 13\% less than the pseudo-slit values. In 17 of
the 31 galaxies with bulges, the pseudo-slit \mlb\ values are 30\%
lower than the map \mlb\ values.  The other 14 galaxies have map \mlb\
values 49\% smaller than the pseudo-slit values.  A few
outliers inflate these averages, so the overall \mld\ differences
are closer to 15\% and the \mlb\ differences are more like 25\%.
Three of the worst outliers in the \mld\ comparison (ESO 381g05, ESO
382g06, ESO 445g39) have quite poor agreement between the pseudo-slit
and velocity map rotation curves.  Two others (ESO 323g39 and ESO
438g08) have slit misalignment issues.  However, the trend for
pseudo-slit \mld\ values to be lower than the map \mld\ values remains.

This trend may have an explanation in the differing methods.  Fitting
pseudo-slit data gives equal weight to each data point regardless of
radial location.  However, velocity map fits utilize more data as
radius increases, giving more weight to the outer regions of the
rotation curves.  Since the outermost points tend to have higher
velocity, the map \mld\ values reflect this by being slightly larger.
We find a similar bias when halos are added -- see below.

As expected for fewer data points, the average pseudo-slit statistical
uncertainties are $\approx 3$ times larger than the velocity map
statistical uncertainties.  The systematic uncertainties due to
position angle uncertainties are of the same magnitude for both
methods, and the estimated systematic error in $\mli \approx 0.5$.  As
with the stars-only map fits, the \mli\ values do not show any trend
with color.

\subsubsection{Halo Fits}

As with the 2-D velocity maps, we have fitted the three halo models to
these rotation curves from 1-D pseudo-slit data.
Figure~\ref{slittomap} shows how the halo mass fraction (within the
radius of the outermost fitted point from the velocity map) compares
between the fits to the pseudo-slit data and the velocity maps.

Not surprisingly, we find constraints on the halo masses are generally
weaker for slit data than for 2-D maps, by a factor $\gtrsim 2$  in
most cases.  The significant outliers in these plots are generally
cases of disagreement between the photometric inclination adopted for
the pseudo-slit, and the kinematic inclination used for the velocity
map.

The slit-estimated halo masses tend to be higher than the
map-estimated values, which we attribute to the above-noted
significance of the large number of pixels in the outer parts of the
map.  Since slit data give greater relative weight to the inner
points, an overfit in the inner parts carries a greater penalty,
sometimes forcing \mld\ to lower values, which requires more halo mass
to match the outer data.

\section{Summary \& Conclusions}

\citet{pw00} present a sample of homogeneously collected, 2-D optical
velocity maps and $I$-band surface brightness distributions of
galaxies.  We have re-analyzed a subsample of these galaxies, omitting
mostly those that are highly inclined, in order to determine whether
such data constrain the dark matter content of these galaxies any more
tightly than do previous studies using longslit data or velocity maps
at lower spatial resolution.

Our results echo those of PW in finding that the stars-only model fits
the shape of the observed rotation curve over a significant radial
range in 34 out of the 40 cases.  This fraction includes 11 cases
where bars or spirals cause a mismatch in the inner barred region
which we argue is acceptable because of non-circular motion.  A
further 5 cases could be fitted tolerably well when the outer data
were discarded, as found by PW.  In addition to these 5 galaxies,
several others have noticeable mass discrepancies in the outer parts.
These stars-only fits yield an average \mld\ of 2.6 and an average
\mlb\ of 2.8, similar to the values found by PW.  The main sources of
uncertainty in these estimates are those arising from distance errors
and possible errors in the adopted projection geometry.

These average \mli\ values, even with no dark matter, are in the range
predicted from stellar population models, but we do not find a trend
of \mli\ with color.  Instead of finding lower \mli\ values for bluer
galaxies, we find a flat distribution which does not decrease for the
bluest galaxies.

We fitted three dark matter halo profiles; two are standard,
pseudo-isothermal (PI) and Navarro-Frenk-White (NFW), and we have also
used a simple power law function over the
radial range of our data.  Three galaxies in our sample do not require
a halo of any type we tried, although these galaxies may have dark
matter halos of some other form.  Should the true inner halo rotation
curve shape resemble that of the luminous matter, any relative
fraction would yield an equally acceptable fit; studies such as this,
which are based on rotation curve data alone, could never break such a
degeneracy.

In most cases, the fit to the data is significantly improved by adding
one or more of our adopted halos, but we do not find a preference for
any particular halo density profile, in the sense that the values of
\biwtm\ are all rather similar no matter which form is adopted.
However, we find wide variations in the \mld\ values and halo masses,
within the radial range spanned by our data, depending upon which halo
density profile is adopted.  This finding implies that even with
high-quality 2-D data, rotation curve fitting is unable to determine
either the disk mass-to-light ratio, halo mass, or density profile
with any degree of confidence.

The pseudo-isothermal and NFW halo models, which have both a length
scale and a density normalization as free parameters, often lead to
fitted \mli\ values that are absurdly low, which seems to indicate
that such functions have too much freedom to adjust their shapes to
quirks in the velocity maps, such as may arise from non-axisymmetric
features in the disk.  These functional forms also lead to strong
degeneracies in the fitted parameters, especially for low halo
densities.  Our suggested power-law halo density profile is more
robust, in the sense that it does not have strong degeneracies and
requires fewer near-massless stellar populations.

Following the suggestion of \citet{alam02}, we have calculated the
central density parameter $\Delta_{v/2}$ for the fitted NFW halos.  We
find that our fitted halo densities are highly uncertain, but seem to
span a very wide range that includes values consistent with the
predictions from cosmological simulations but also extend to much lower
densities; a few galaxies have no detectable dark matter at all.

Fits with the NFW halo function generally attribute the least mass to
the disk, and the most to the halo, while the power law prefers the
highest disk masses.  On average, the power-law halos have radial
density profiles $\propto r^{-1}$ or somewhat steeper, \ie\ a slope
resembling the inner parts of halo density profiles from recent
cosmological simulations of structure formation \citep{dms04}.
However, the corresponding mean halo densities are generally lower
than those obtained with the NFW function.

\begin{figure}
\epsscale{0.8}
\plotone{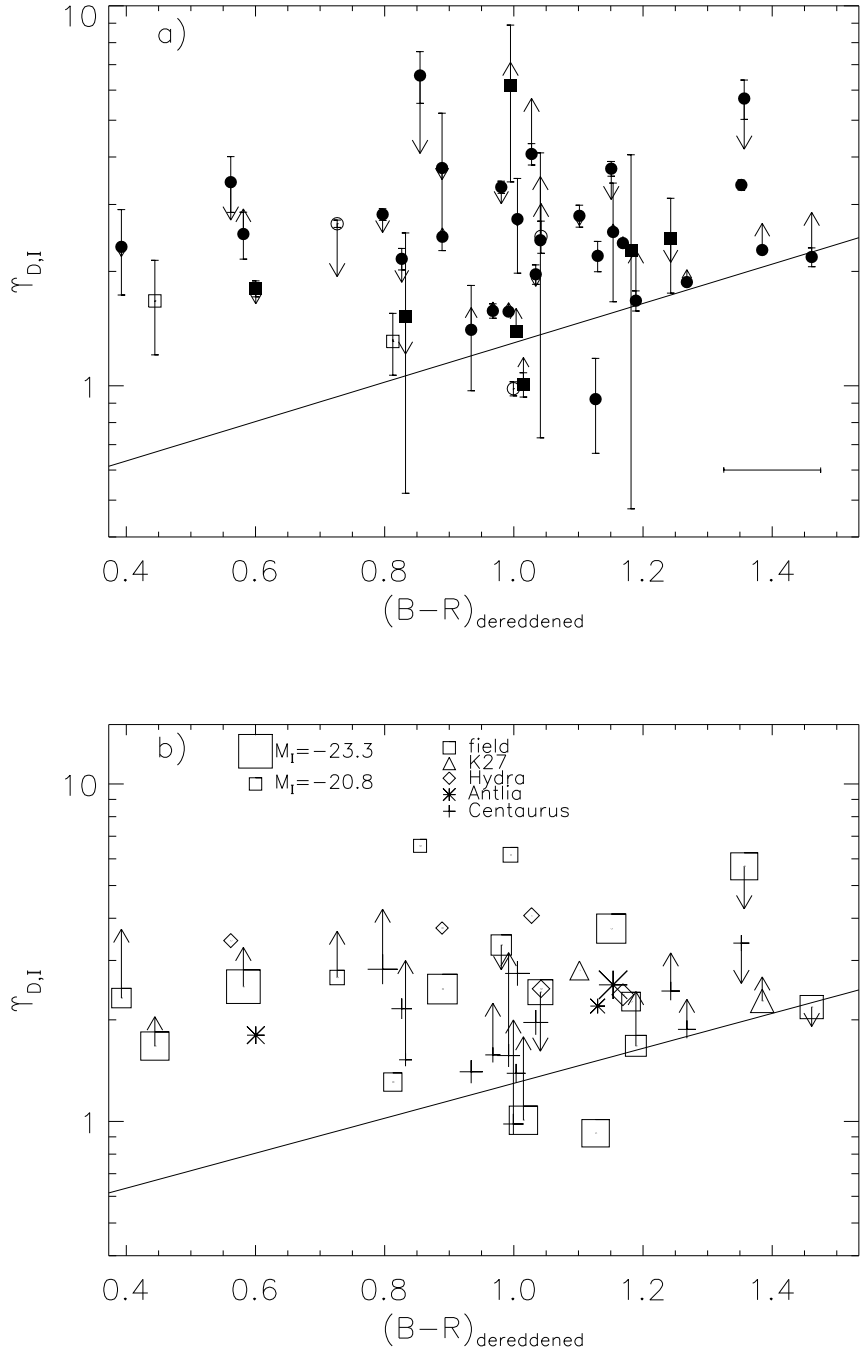}
\caption{Fitted \mld\ values for stars-only vs.\
dereddened $B-R$.  In both panels the line is the best-fit correlation
found by \citet{bell03}.  (a) Estimated systematic errors are shown
and the arrows indicate how the points would move were the photometric
inclinations used to derive \mld\ values.  Squares denote galaxies
without bulges.  Galaxies with evidence of mass discrepancies are
marked by open symbols, and their \mld\ values are the maximum-disk
values from Table \ref{btab}.  The horizontal errorbar in the lower
right corner indicates the estimated error in the color adopted for
each galaxy.  (b) The same plot, but the different symbols denote
whether a galaxy belongs to a cluster or the field.  Larger symbol
sizes correspond to more luminous (and more metal rich) galaxies.  The
arrows show how points would move if their Tully-Fisher distances were
used instead of Hubble distances.  \label{barycolplot}}
\end{figure}

\begin{figure}
\plotone{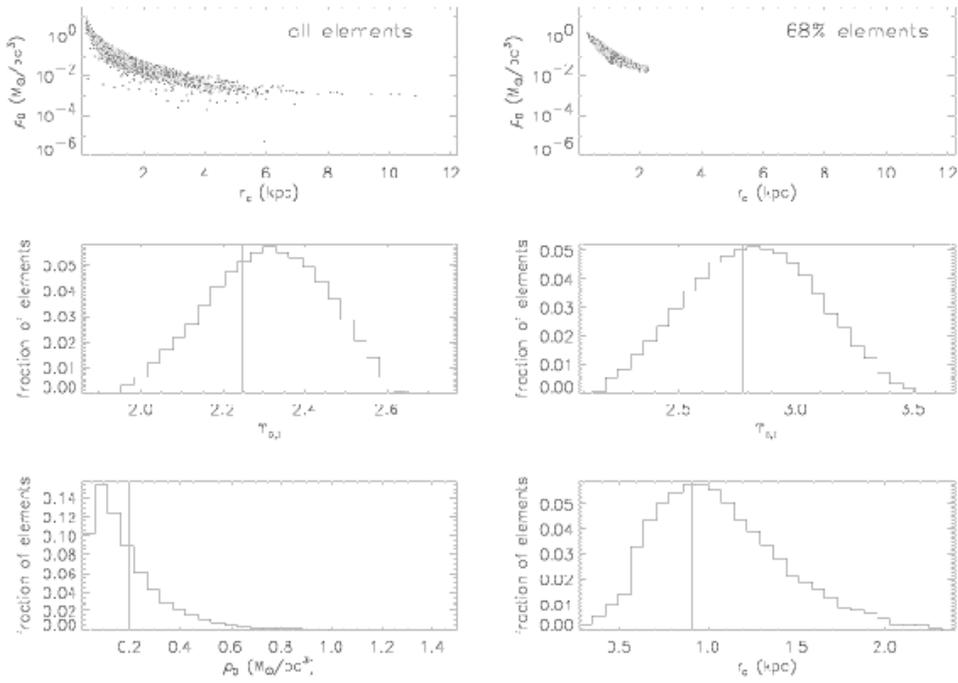}
\caption{Top left shows the projection of all
Markov Chain Monte Carlo elements onto the $\rho_0$-$v_{\infty}$ plane
for the PI halo fit to ESO 322g82, to illustrate the strong degeneracy
of these parameters.  (Only 1 point in 300 is plotted.)  Top right
shows the same projection of the 68\% of points lying in the densest
part of the parameter hypervolume.  The other four panels show
histograms of the same 68\% of selected points projected onto the four
parameter axes.  The vertical line in each panel shows the best fit
value. \label{mchist}}
\end{figure}

\begin{figure}
\plotone{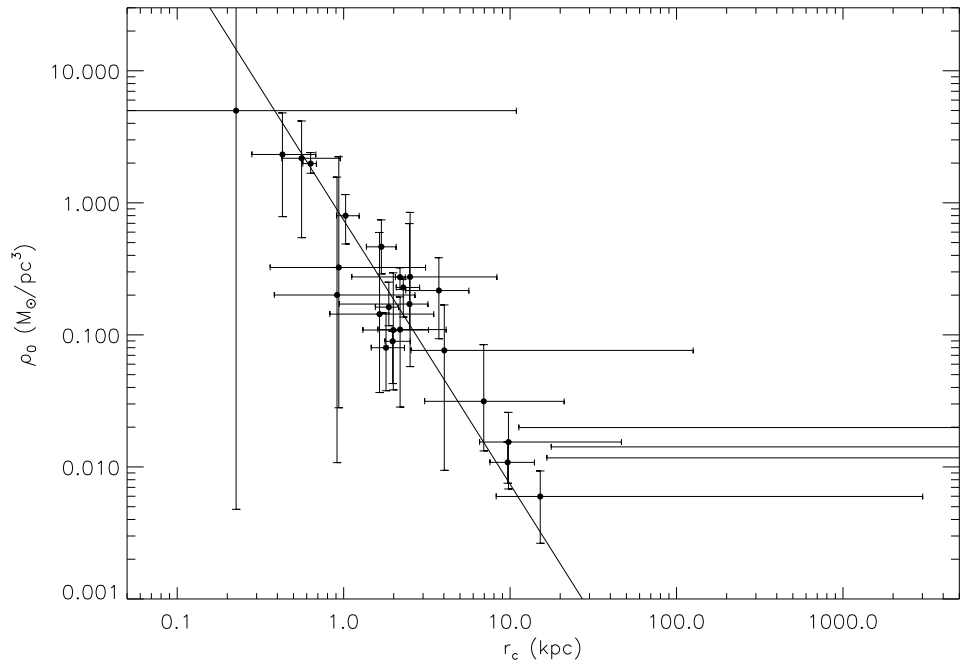}
\caption{Central density versus scale radius for
the PI fits.  The solid line is not a fit to the points, but shows the
relation between the parameters for $v_{\infty}=200$ km/s.  Note a
marginal tendency for the halos with higher central densities to fall
below the line while lower density halos lie above it. \label{pivals}}
\end{figure}

We have compared our fitted \mld\ values with population synthesis
models.  While our stars-only \mld\ values (Fig.\ \ref{barycolplot})
are in the range predicted by \citet{bell03}, we do not find a trend
with color.  When halos are included (Fig.\ \ref{halocolplot}), many
\mld\ values are lower than their predictions.  Distance errors affect
the estimated \mli\ directly, but it is unlikely they could be large
enough to bring the points into agreement with the prediction;
frequently, adopting Tully-Fisher distances instead of Hubble distances
would reduce \mld\ still further.  Also, we do not see a
correlation between color and distance, luminosity (metallicity), or
cluster membership needed to recover the predicted trend.  The most
likely explanation is that none of our models has captured the actual
distribution of dark matter in our sample of galaxies and, without
this knowlegde, our fits cannot determine the true \mld.

We have compared our results from fitting 2-D velocity maps with those
from pseudo-slit data extracted from our data on the photometric major
axis only.  Derived rotation curves generally agree reasonably well
with the rotation curves from the velocity maps.  In a few cases,
disagreement stems from misaligned slits, poor major axis data, and/or
nonaxisymmetric structure.  Not surprisingly, we find the
uncertainties in the estimated parameters are $\approx 3$ times larger
from pseudo-slit data than from the 2-D maps.  We find a tendency for
\mli\ values from our pseudo-slit data to be $\approx 15\%$ lower than
the corresponding velocity map values, which we attribute to the extra
weight that 2-D maps give to the larger number of data values in the
outer parts of the map.  The greatest improvements afforded by 2-D
velocity maps is that they offer tighter constraints on the projection
geometry and the systemic velocity, as well as revealing more
extensive rotation curves in cases with sparse data along their major
axes.

Fits of axisymmetric models to two-dimensional
velocity maps, even at optical resolution, do not provide significant
constraints on either the mass-to-light of the luminous matter or the
halo mass fraction or its density profile.  Additional data extending
to larger radii, \eg\ from neutral hydrogen maps, must tighten the
constraints somewhat, but the improvement is not dramatic; \eg\
\citet{bac01} still find large variations in \mld\ and halo mass
between the different halo models.  We therefore conclude that
additional information is needed to constrain the disk mass, and
therefore also the halo density, in a galaxy.

\begin{figure}
\plotone{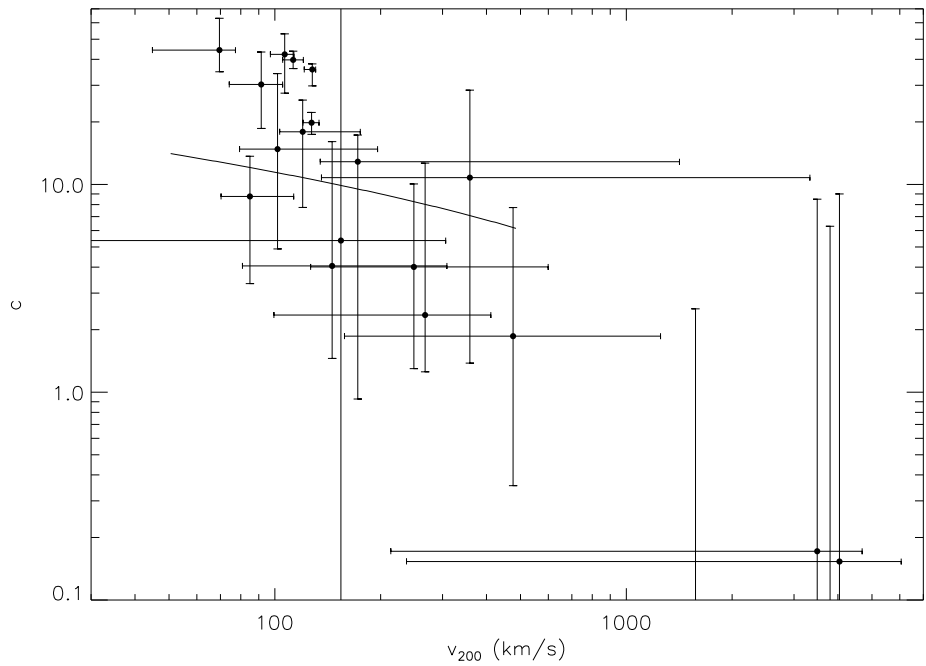}
\caption{Concentration versus $v_{200}$ for the
NFW fits.  Note how the uncertainties rapidly increase as the halos
become less dense and more extended.  The solid line illustrates the
mean correlation found in the simulation of \citet{bull01}.
\label{nfwvals}}
\end{figure}

\begin{figure}
\plotone{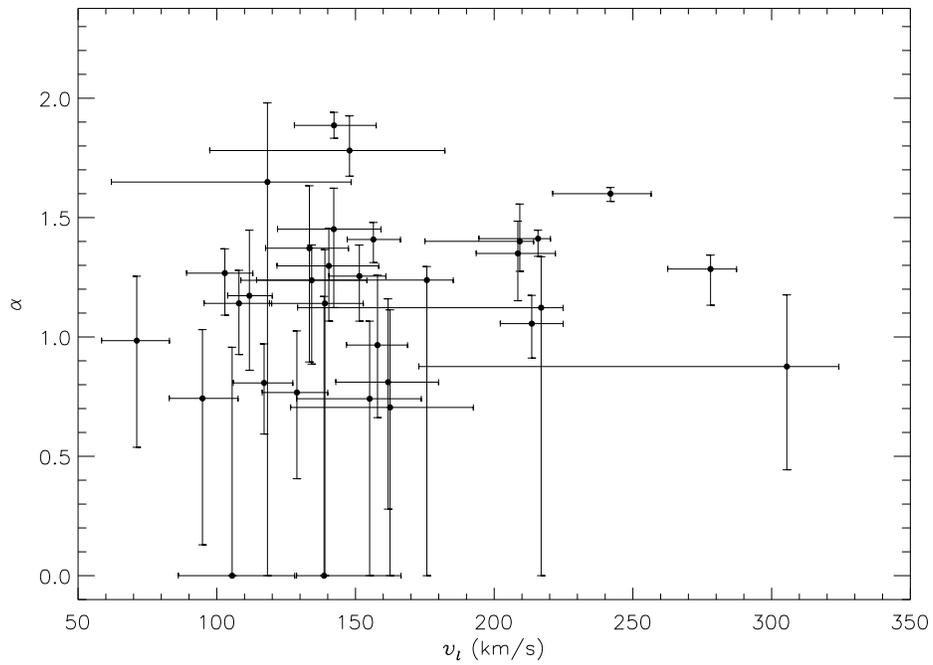}
\caption{Slope parameter $\alpha$ versus $v_l$ for
the power-law fits.  The $\alpha$ values have an average value of 1.1,
but have a wide range. \label{powvals}}
\end{figure}

\begin{figure}
\plotone{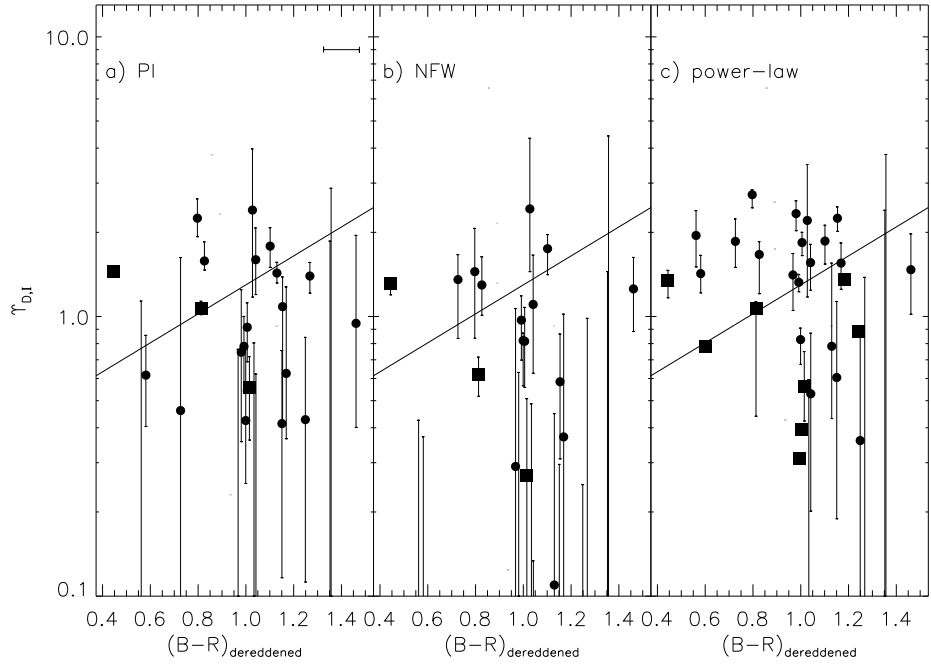}
\caption{Fitted \mld\ values vs. dereddened $B-R$
when halos are included.  The solid line in each panel is the best-fit
correlation found by \citet{bell03}.  As in Figure
\ref{barycolplot}(a), the solid squares denote galaxies without
bulges.  The error bars shown are the statistical errors presented in
the tables, and do not include systematic errors.  The horizontal
error bar in the upper right corner of panel a) is the estimated color
uncertainty for each galaxy.
\label{halocolplot}}
\end{figure}

\begin{figure}
\plotone{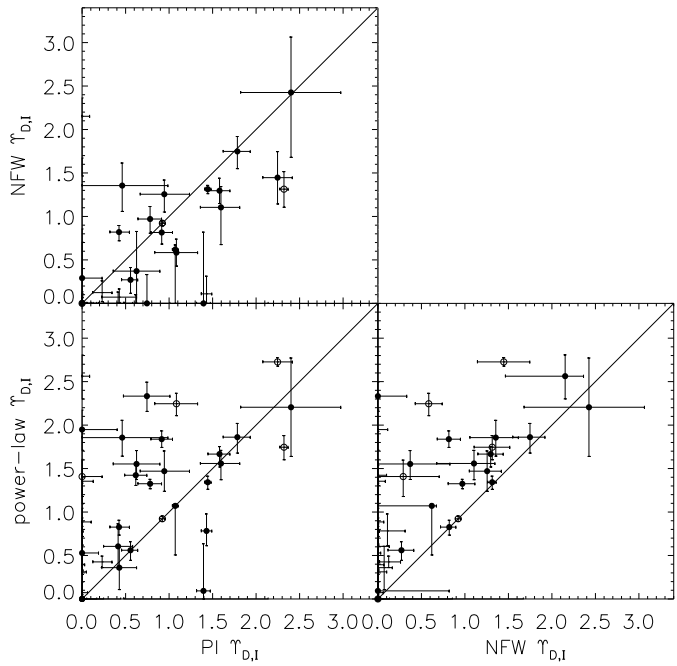}
\caption{Comparisons between the \mld\ values
fitted for the different halo models.  Filled points mark galaxies
which have $\mld > 0.5$ for both of the relevant models.  Open circles
indicate galaxies which do not require halos for either of the models.
\label{mlcompare}}
\end{figure}

\begin{figure}
\plotone{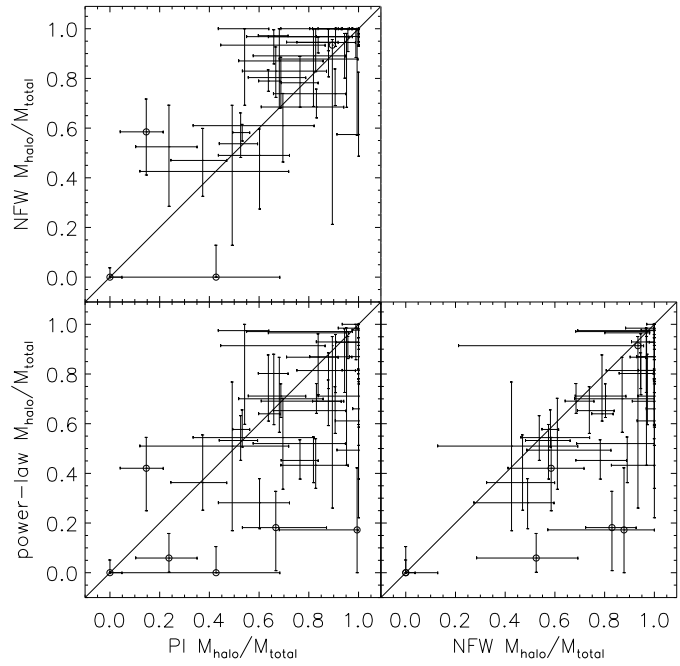}
\caption{Comparisons between the fractional halo
masses for the different halo models.  Open and filled points have the
same meanings as in Figure \ref{mlcompare}. \label{maptomap}}
\end{figure}

\begin{figure}
\plotone{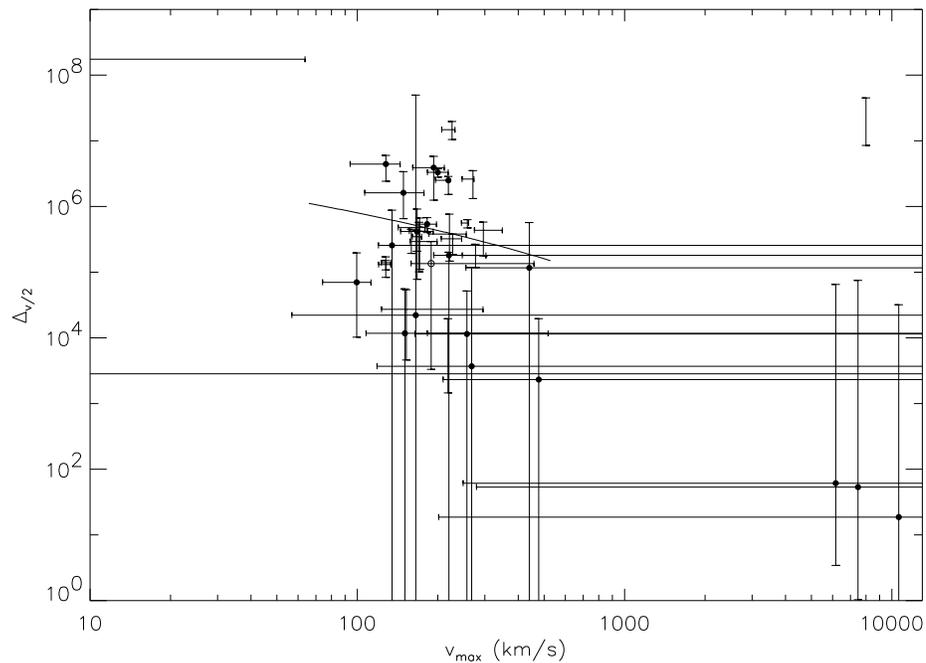}
\caption{Inner halo densities ($\Delta_{v/2}$) as
a function of maximum halo velocity for the NFW halo function.  Filled
circles denote galaxies which have best-fit $\mld>0.5$, the open
circle is for a galaxy for which no halo is required, and the points
with error bars, but no circle are those for which $\mld<0.5$.  The
large error bars are for galaxies for which $r_s$ is poorly
constrained by our data.  The solid line is the predicted correlation
from \citet{zent02} and the error bar in the upper right hand corner
is the $\Delta_{v/2}$ uncertainty estimated from \citet{bull01}.
\label{simcomp}}
\end{figure}

\begin{figure}
\plotone{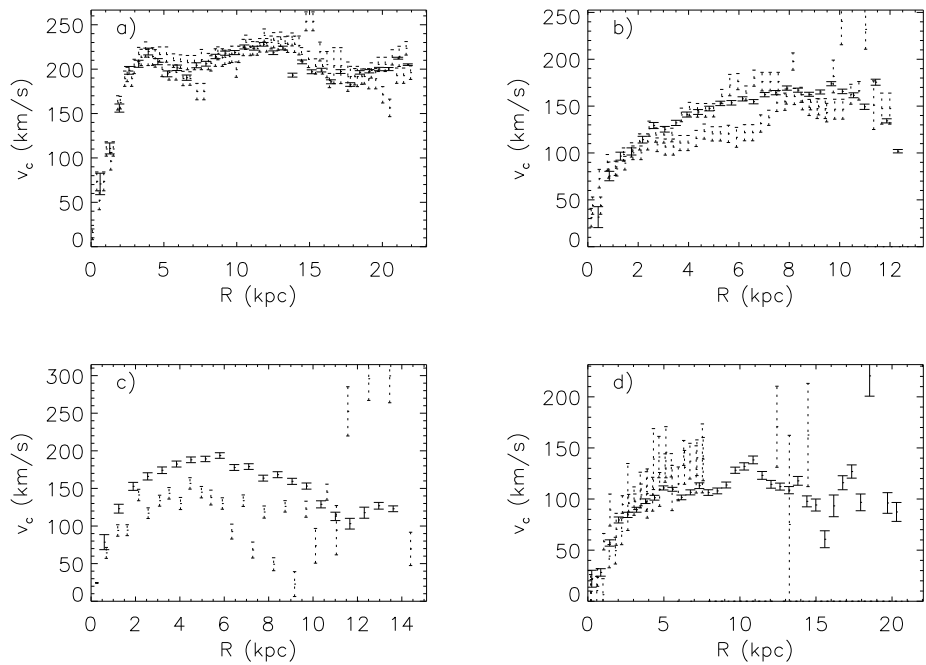}
\caption{Comparisons between velocity map
rotation curves (solid error bars) and pseudo-slit rotation curves
(dotted error bars) for 4 galaxies.  (a) ESO 322g82 is an example of
good agreement between the rotation curves.  (b) ESO 322g36 has strong
spirals which cause a difference between the approaching and receding
sides of the pseudo-slit data.  (c) ESO 438g08, the poor
correspondence between the two curves arises because the photometric
major axis is misaligned from the kinematic major axis.  (d) ESO
501g01: the data along the major axis simply does not extend to large
radius. \label{rccomp}}
\end{figure}

\begin{figure}
\plotone{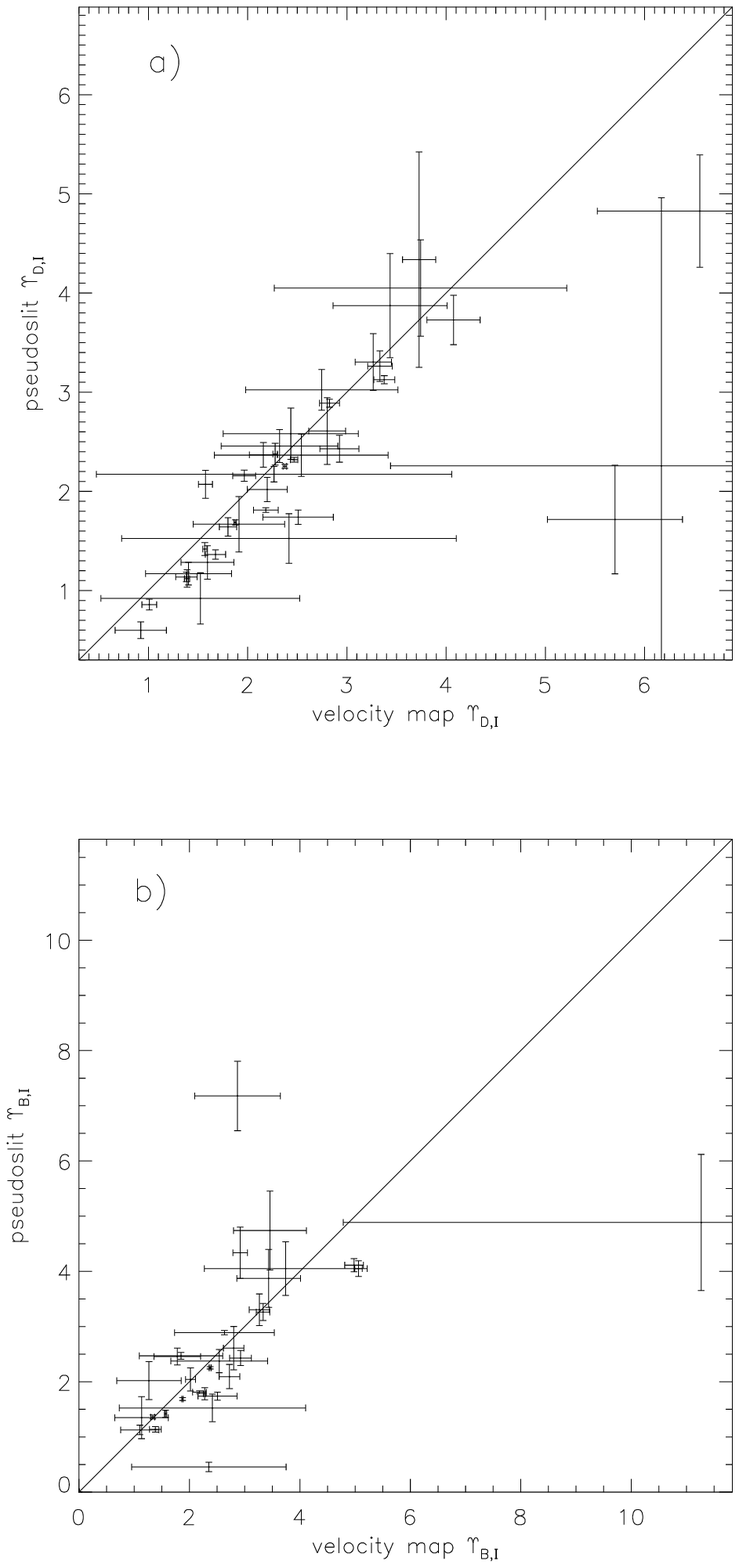}
\caption{Comparisons between stars-only velocity
map \mli\ values and pseudo-slit \mli\ values. \label{rcmlcomp}}
\end{figure}

\begin{figure}
\plotone{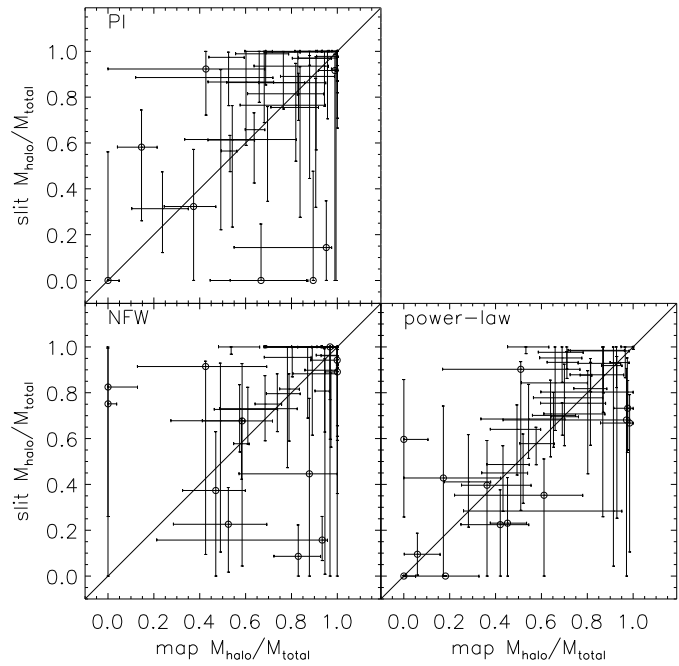}
\caption{Comparisons of the dark matter halo
masses for each of the halo models derived using pseudo-slit versus
map data.  Open circles indicate galaxies which do not require halos
for either the pseudoslit or velocity map fit.  \label{slittomap}}
\end{figure}

\acknowledgements
The authors gratefully acknowledge the assistance and advice of
Povilas Palunas, Ted Williams, Tad Pryor, and Stacy McGaugh.  We thank
Eric Peng, Pat C\^ot\'e, and Laura Ferrarese for a number of helpful
discussions.  We also acknowledge James Bullock for supplying us with
the cosmological simulation results shown in Figures \ref{nfwvals} and
\ref{simcomp}.  An anonymous referee supplied a detailed report based
on a very careful read that helped us to improve the paper.  This work
has been supported by NASA grant NAG 5-10110.  AK is a Cottrell
Scholar of the Research Corporation.

\onecolumn

\clearpage

\clearpage

\clearpage

\clearpage

\clearpage

\clearpage

\clearpage

\clearpage

\clearpage

\clearpage

\clearpage

\clearpage

\clearpage

\clearpage

\begin{deluxetable}{lll} 
\tablewidth{0pt} 
\tablecaption{\citet{pw00} galaxies rejected for this study and
reasons for exclusion.\label{omitted}}
\tablehead{\colhead{Galaxy} & 
\colhead{Galaxy} & \colhead{Galaxy}} 
\startdata 
Abell 1644d83\tablenotemark{1} & ESO 377g11$^3$ & ESO 444g21$^1$ \\  
ESO 216g20\tablenotemark{2} & ESO 381g51$^1$ & ESO 444g86$^1$ \\
ESO 267g30$^1$ & ESO 382g58$^{1,3}$ & ESO 445g35$^3$ \\
ESO 269g61$^1$ & ESO 383g02$^3$ & ESO 445g58$^{2,3}$ \\
ESO 322g19$^1$ & ESO 383g88$^3$ & ESO 445g81$^1$ \\
ESO 322g48$^1$ & ESO 435g50$^1$ & ESO 501g11$^1$ \\
ESO 322g87$^1$ & ESO 436g39$^1$ & ESO 501g15$^{2,3}$ \\
ESO 374g02\tablenotemark{3} & ESO 437g30$^1$ & ESO 501g86$^2$ \\
ESO 375g12$^3$ & ESO 437g34$^2$ & ESO 509g91$^1$ \\
ESO 375g29$^1$ & ESO 437g54$^1$ & ESO 510g11$^3$ \\
ESO 376g02$^1$ & ESO 441g22$^3$ & ESO 572g17$^3$ \\
ESO 376g10$^{1,2,3}$ & & \\
\enddata 
\tablenotetext{1}{\hspace{0.1cm}inclination $> 75^{\circ}$}
\tablenotetext{2}{\hspace{0.1cm}sparse H$\alpha$ emission}
\tablenotetext{3}{\hspace{0.1cm}lack of central H$\alpha$ emission}
\end{deluxetable} 

\begin{deluxetable}{ccccccc} 
\tablewidth{0pt} 
\tablecaption{Best-fit \biwtm values, parameter
values, and uncertainties for stars-only models.\label{btab}} 
\tablehead{\colhead{\#} & \colhead{Galaxy} & \colhead{\biwtm} & 
\colhead{\mld (M$_{\sun}$/L$_{\sun}$)} & 
\colhead{$\sigma_{\mld}$} & 
\colhead{\mlb (M$_{\sun}$/L$_{\sun}$)} & 
\colhead{$\sigma_{\mlb}$}} 
\startdata 
1 & ESO 215g39 & 1.5836 & 2.32 & 0.59 & 
1.78 & 0.42 \\  
2 & ESO 216g20 & 4.4338 & 2.18 & 0.12 & 
2.18 & 0.12 \\  
3 & ESO 263g14 & 2.1086 & 1.01 & 0.07 & 
\nodata & \nodata \\  
4 & ESO 267g29 & 5.6170 & 2.42 & 1.69 & 
2.42 & 1.69 \\  
5 & ESO 268g37 & 4.8726 & 2.43 & 0.68 & 
\nodata & \nodata \\  
6 & ESO 268g44 & 2.0330 & 1.57 & 0.07 & 
2.73 & 0.19 \\  
7 & ESO 317g41 & 4.5788 & 2.47 & 0.04 & 
1.34 & 0.03 \\  
8 & ESO 322g36/NGC 4575 & 4.9430 & 1.40 & 0.43 & 
1.10 & 0.35 \\  
9 & ESO 322g42 & 4.1576 & 1.39 & 0.03 & 
\nodata & \nodata \\  
10 & ESO 322g44 & 2.4256 & 1.38 & 0.11 & 
1.38 & 0.11 \\  
11 & ESO 322g45 & 2.5188 & 1.87 & 0.01 & 
1.87 & 0.01 \\  
12 & ESO 322g76 & 2.0912 & 2.16 & 0.14 & 
1.13 & 0.48 \\  
13 & ESO 322g77/NGC 4696a & 3.3364 & 3.38 & 0.11 & 
1.85 & 0.76 \\  
14 & ESO 322g82/NGC 4679 & 4.0750 & 2.82 & 0.10 & 
2.63 & 0.90 \\  
15 & ESO 323g25 & 2.7170 & 2.75 & 0.77 & 
11.27 & 6.48 \\  
16 & ESO 323g27 & 2.8732 & 1.96 & 0.12 & 
5.06 & 0.07 \\  
17 & ESO 323g39 & 5.1232 & 1.52 & 1.00 & 
\nodata & \nodata \\  
18 & ESO 323g42 & 1.5962 & 1.57 & 0.02 & 
1.57 & 0.02 \\  
19 & ESO 323g73 & 3.6364 & 1.59 & 0.27 & 
\nodata & \nodata \\  
20 & ESO 374g03 & 3.4210 & 1.80 & 0.09 & 
\nodata & \nodata \\  
21 & ESO 375g02/IC 2559 & 2.2998 & 2.20 & 0.20 & 
1.27 & 0.58 \\  
22 & ESO 381g05 & 3.0762 & 6.17 & 2.73 & 
\nodata & \nodata \\  
23 & ESO 382g06 & 2.4402 & 6.56 & 1.03 & 
3.46 & 0.66 \\  
24 & ESO 435g26/NGC 3095 & 11.8916 & 2.54 & 0.88 & 
2.54 & 0.88 \\  
25 & ESO 437g04 & 3.2496 & 2.92 & 0.20 & 
2.92 & 0.20 \\  
26 & ESO 437g31 & 1.2074 & 3.74 & 1.47 & 
3.74 & 1.47 \\  
27 & ESO 438g08 & 2.0276 & 0.92 & 0.26 & 
2.35 & 1.40 \\  
28 & ESO 438g15 & 2.7118 & 1.67 & 0.10 & 
2.02 & 0.09 \\  
29 & ESO 439g18 & 4.5172 & 2.51 & 0.35 & 
2.51 & 0.35 \\  
30 & ESO 439g20 & 2.0360 & 3.33 & 0.12 & 
3.33 & 0.12 \\  
31 & ESO 444g47 & 2.0842 & 3.27 & 0.18 & 
3.27 & 0.18 \\  
32 & ESO 445g15 & 3.3342 & 2.80 & 0.19 & 
2.80 & 0.19 \\  
33 & ESO 445g19/IC 4319 & 2.4596 & 2.28 & 0.02 & 
2.28 & 0.02 \\  
34 & ESO 445g39/NGC 5298 & 4.2358 & 5.70 & 0.68 & 
2.87 & 0.78 \\  
35 & ESO 446g01 & 2.6448 & 1.91 & 0.46 & 
\nodata & \nodata \\  
36 & ESO 501g01 & 1.9834 & 3.34 & 0.72 & 
3.34 & 0.72 \\  
37 & ESO 501g68 & 2.1836 & 4.07 & 0.27 & 
4.98 & 0.17 \\  
38 & ESO 502g02/NGC 3463 & 1.9144 & 2.37 & 0.02 & 
2.37 & 0.02 \\  
39 & ESO 509g80/IC 4298 & 3.6484 & 3.73 & 0.17 & 
2.92 & 0.13 \\  
40 & ESO 569g17/NGC 3453 & 1.6434 & 2.26 & 1.79 & 
\nodata & \nodata \\  
\cutinhead{Maximum allowed \mli\ values from fits with the outer
data discarded.}
10 & ESO 322g44 & 2.1920 & 0.98 & 0.04 & 
0.98 & 0.04 \\  
19 & ESO 323g73 & 1.1802 & 1.31 & 0.24 & 
\nodata & \nodata \\  
25 & ESO 437g04 & 3.2294 & 2.47 & 0.24 & 
2.47 & 0.24 \\  
31 & ESO 444g47 & 1.2608 & 2.67 & 0.06 & 
2.67 & 0.06 \\  
35 & ESO 446g01 & 1.1288 & 1.67 & 0.47 & 
\nodata & \nodata \\  
\enddata 
\end{deluxetable} 

\begin{deluxetable}{lcccccc} 
\tablewidth{0pt} 
\tablecaption{Best-fit $b^2$ values, parameter values,
and uncertainties for PI models.\label{itab}} 
\tablehead{\colhead{Galaxy} & 
\colhead{$b^2$} & 
\colhead{\mld (M$_{\sun}$/L$_{\sun}$)} & 
\colhead{\mlb (M$_{\sun}$/L$_{\sun}$)} & 
\colhead{$\rho_0$ (M$_{\sun}$/pc$^3$)} & 
\colhead{r$_c$ (kpc)} & 
\colhead{$v_{\infty}$ (km/s)}\tablenotemark{a}} 
\startdata 
1 & 1.3822 & 
1.5e-06 $^{0.36}_{1.3e-06}$ &  
1.50 $^{1.50}_{1.50}$ &  
0.21 $^{0.28}_{0.12}$ &  
1.7 $^{1.1}_{0.6}$ &  
176.46 \\  
2 & 4.0878 & 
0.95 $^{1.01}_{0.54}$ &  
2.12 $^{0.47}_{0.58}$ &  
7.6e-02 $^{9.2e-02}_{6.7e-02}$ &  
4.0 $^{121.8}_{1.5}$ &  
258.44 \\  
3 & 1.8988 & 
0.56 $^{0.16}_{0.19}$ &  
\nodata $^{ }_{ }$ &  
0.11 $^{0.19}_{7.0e-02}$ &  
2.0 $^{1.2}_{0.7}$ &  
152.39 \\  
4 & 5.3118 & 
1.60 $^{0.48}_{0.40}$ &  
1.95 $^{0.78}_{1.01}$ &  
3.1e-02 $^{5.3e-02}_{1.8e-02}$ &  
6.9 $^{14.2}_{3.9}$ &  
286.14 \\  
5 & 3.9596 & 
9.2e-07 $^{0.29}_{8.8e-07}$ &  
\nodata $^{ }_{ }$ &  
0.20 $^{3.7e-02}_{7.0e-02}$ &  
1.7 $^{0.3}_{0.1}$ &  
181.35 \\  
6 & 1.9840 & 
1.2e-08 $^{0.76}_{1.2e-08}$ &  
0.28 $^{1.45}_{0.28}$ &  
2.32 $^{2.48}_{1.54}$ &  
0.4 $^{0.3}_{0.1}$ &  
151.97 \\  
7 & 4.5604\tablenotemark{*} & 
2.32 $^{0.10}_{9.1e-02}$ &  
1.43 $^{0.18}_{0.21}$ &  
1.9e-03 $^{9.9e-04}_{1.2e-03}$ &  
1.2e+04 $^{0.8}_{11.7}$ &  
1.2e+05 \\  
8 & 4.4864 & 
0.23 $^{0.27}_{0.23}$ &  
0.77 $^{0.33}_{0.34}$ &  
0.13 $^{0.11}_{6.9e-02}$ &  
2.2 $^{1.0}_{0.6}$ &  
185.04 \\  
9 & 3.3828 & 
1.1e-07 $^{0.10}_{5.7e-08}$ &  
\nodata $^{ }_{ }$ &  
2.9e-02 $^{4.2e-03}_{5.6e-03}$ &  
4.1 $^{0.6}_{0.3}$ &  
161.18 \\  
10 & 1.5610 & 
0.42 $^{0.36}_{0.17}$ &  
0.42 $^{0.36}_{0.17}$ &  
0.32 $^{1.91}_{0.30}$ &  
0.9 $^{2.2}_{0.6}$ &  
123.76 \\  
11 & 2.0248 & 
1.40 $^{0.17}_{0.18}$ &  
0.90 $^{0.90}_{0.75}$ &  
1.2e-02 $^{3.4e-03}_{3.2e-03}$ &  
2.6e+04 $^{1.2e+06}_{2.6e+04}$ &  
6.5e+05 \\  
12 & 2.0442 & 
1.58 $^{0.27}_{0.11}$ &  
1.69 $^{0.22}_{0.32}$ &  
6.0e-03 $^{3.3e-03}_{3.3e-03}$ &  
15.2 $^{3.0e+03}_{6.9}$ &  
272.67 \\  
13 & 3.0918 & 
2.0e-05 $^{1.86}_{1.9e-05}$ &  
4.19 $^{0.81}_{2.26}$ &  
0.27 $^{0.57}_{0.22}$ &  
2.5 $^{5.8}_{1.4}$ &  
305.55 \\  
14 & 4.0594 & 
2.25 $^{0.39}_{0.32}$ &  
2.77 $^{0.79}_{0.65}$ &  
0.20 $^{1.37}_{0.19}$ &  
0.9 $^{1.8}_{0.5}$ &  
94.99 \\  
15 & 2.5246 & 
0.92 $^{0.20}_{0.23}$ &  
0.92 $^{0.20}_{0.23}$ &  
1.98 $^{0.42}_{0.31}$ &  
0.6 $^{5.3e-02}_{6.5e-02}$ &  
207.34 \\  
16 & 2.6860 & 
8.0e-07 $^{0.80}_{8.0e-07}$ &  
2.89 $^{0.96}_{1.44}$ &  
0.17 $^{0.53}_{6.1e-02}$ &  
2.5 $^{0.7}_{1.5}$ &  
239.32 \\  
17 & 4.3368 & 
7.6e-08 $^{0.17}_{7.6e-08}$ &  
\nodata $^{ }_{ }$ &  
4.0e-02 $^{1.4e-02}_{1.5e-02}$ &  
2.9 $^{0.9}_{0.5}$ &  
136.76 \\  
18 & 1.5142 & 
0.78 $^{0.22}_{0.24}$ &  
0.78 $^{0.22}_{0.24}$ &  
8.0e-02 $^{6.7e-02}_{4.2e-02}$ &  
1.8 $^{0.5}_{0.3}$ &  
118.18 \\  
19 & 2.8006 & 
1.07 $^{6.3e-02}_{5.6e-02}$ &  
\nodata $^{ }_{ }$ &  
2.0e-02 $^{2.2e-03}_{1.8e-03}$ &  
1.7e+04 $^{4.4e+05}_{1.7e+04}$ &  
5.6e+05 \\  
20 & 3.1678 & 
1.6e-07 $^{3.9e-02}_{1.5e-07}$ &  
\nodata $^{ }_{ }$ &  
0.10 $^{1.4e-02}_{1.3e-02}$ &  
1.8 $^{0.2}_{0.1}$ &  
135.16 \\  
21 & 2.0276 & 
1.43 $^{0.13}_{0.11}$ &  
2.62 $^{0.52}_{0.47}$ &  
1.4e-02 $^{2.1e-03}_{1.9e-03}$ &  
3.9e+04 $^{2.4e+06}_{3.9e+04}$ &  
1.1e+06 \\  
22 & 2.2082 & 
5.9e-08 $^{0.26}_{2.2e-08}$ &  
\nodata $^{ }_{ }$ &  
0.50 $^{0.11}_{9.1e-02}$ &  
1.7 $^{0.2}_{0.2}$ &  
287.02 \\  
23 & 2.4352\tablenotemark{*} & 
3.79 $^{3.36}_{2.89}$ &  
1.94 $^{1.95}_{1.42}$ &  
0.11 $^{1.3e+04}_{0.11}$ &  
1.5 $^{5.5}_{1.5}$ &  
117.94 \\  
24 & 11.8492 & 
1.08 $^{0.30}_{0.67}$ &  
1.51 $^{1.41}_{0.89}$ &  
0.16 $^{8.8e-02}_{4.5e-02}$ &  
1.9 $^{0.3}_{0.3}$ &  
174.89 \\  
25 & 2.4646 & 
1.7e-06 $^{0.62}_{1.7e-06}$ &  
9.8e-07 $^{0.18}_{9.7e-07}$ &  
0.23 $^{3.6e-02}_{9.2e-02}$ &  
2.3 $^{0.6}_{0.2}$ &  
253.65 \\  
26 & 1.0336 & 
1.8e-08 $^{0.34}_{1.8e-08}$ &  
1.8e-08 $^{0.34}_{1.8e-08}$ &  
0.19 $^{3.2e-02}_{4.1e-02}$ &  
1.5 $^{0.1}_{0.1}$ &  
151.99 \\  
27 & 2.0276\tablenotemark{*} & 
0.92 $^{5.7e-02}_{6.6e-02}$ &  
2.35 $^{0.61}_{0.63}$ &  
3.6e-07 $^{2.0e+09}_{3.6e-07}$ &  
9.4e-02 $^{5.6}_{9.4e-02}$ &  
0.01 \\  
28 & 2.5456 & 
0.43 $^{0.41}_{0.32}$ &  
1.92 $^{0.24}_{0.18}$ &  
1.5e-02 $^{1.0e-02}_{8.6e-03}$ &  
9.8 $^{36.9}_{3.2}$ &  
282.29 \\  
29 & 3.8966 & 
0.62 $^{0.24}_{0.21}$ &  
0.62 $^{0.24}_{0.21}$ &  
0.46 $^{0.28}_{0.17}$ &  
1.7 $^{0.4}_{0.3}$ &  
266.99 \\  
30 & 1.7666 & 
0.75 $^{0.50}_{0.39}$ &  
0.75 $^{0.50}_{0.39}$ &  
0.80 $^{0.36}_{0.31}$ &  
1.0 $^{0.2}_{0.1}$ &  
213.50 \\  
31 & 1.6208 & 
0.46 $^{1.16}_{0.46}$ &  
0.46 $^{1.16}_{0.46}$ &  
0.11 $^{8.4e-02}_{8.1e-02}$ &  
2.2 $^{1.9}_{0.6}$ &  
168.11 \\  
32 & 3.0656 & 
1.78 $^{0.29}_{0.29}$ &  
1.78 $^{0.29}_{0.29}$ &  
0.14 $^{0.45}_{0.11}$ &  
1.6 $^{1.8}_{0.8}$ &  
144.76 \\  
33 & 2.2896 & 
2.9e-06 $^{0.32}_{2.8e-06}$ &  
1.55 $^{0.73}_{0.72}$ &  
0.19 $^{5.0e-02}_{5.1e-02}$ &  
2.4 $^{0.4}_{0.3}$ &  
241.52 \\  
34 & 4.1728 & 
3.3e-06 $^{2.88}_{3.3e-06}$ &  
3.29 $^{1.09}_{1.38}$ &  
0.22 $^{0.17}_{0.12}$ &  
3.7 $^{1.9}_{1.4}$ &  
404.20 \\  
35 & 2.2198 & 
1.44 $^{5.9e-02}_{6.0e-02}$ &  
\nodata $^{ }_{ }$ &  
1.1e-02 $^{4.6e-03}_{3.3e-03}$ &  
9.7 $^{4.3}_{2.1}$ &  
233.82 \\  
36 & 2.2832 & 
7.2e-10 $^{1.14}_{2.9e-18}$ &  
6.7e-07 $^{0.64}_{4.4e-07}$ &  
8.9e-02 $^{1.7e-02}_{4.7e-02}$ &  
2.0 $^{0.5}_{0.2}$ &  
136.94 \\  
37 & 2.1342 & 
2.40 $^{1.57}_{1.23}$ &  
2.79 $^{2.19}_{1.33}$ &  
4.99 $^{3.4e+07}_{4.99}$ &  
0.2 $^{10.7}_{0.2}$ &  
117.32 \\  
38 & 1.6984 & 
0.63 $^{0.65}_{0.26}$ &  
0.63 $^{0.65}_{0.26}$ &  
2.18 $^{1.99}_{1.63}$ &  
0.6 $^{0.4}_{0.1}$ &  
191.69 \\  
39 & 3.4298 & 
0.41 $^{0.34}_{0.30}$ &  
0.41 $^{0.34}_{0.30}$ &  
0.27 $^{4.9e-02}_{5.3e-02}$ &  
2.2 $^{0.2}_{0.1}$ &  
265.12 \\  
40 & 1.4480 & 
6.1e-10 $^{0.30}_{6.1e-10}$ &  
\nodata $^{ }_{ }$ &  
8.24 $^{2.88}_{2.05}$ &  
0.4 $^{4.9e-02}_{5.4e-02}$ &  
243.78 \\  
\enddata 
\tablenotetext{a}{$v_{\infty}$ is not a fitted parameter.
It has been calculated using the best-fit $\rho_0$ and r$_c$ 
values to facilitate comparison with the other halos' velocity
parameters.} 
\tablenotetext{*}{Indicates that the halo fit is 
no better than the stars-only fit at the 95\% level.} 
\end{deluxetable} 

\begin{deluxetable}{lcccccc} 
\tablewidth{0pt} 
\tablecaption{Best-fit $b^2$ values, parameter 
values, and uncertainties for NFW models.\label{ntab}} 
\tablehead{\colhead{Galaxy} & 
\colhead{$b^2$} & 
\colhead{\mld (M$_{\sun}$/L$_{\sun}$)} & 
\colhead{\mlb (M$_{\sun}$/L$_{\sun}$)} & 
\colhead{c} & 
\colhead{$v_{200}$ (km/s)} & 
\colhead{r$_s$ (kpc)}\tablenotemark{a}} 
\startdata 
1 & 1.3946 & 
3.9e-07 $^{0.15}_{3.3e-07}$ &  
1.13 $^{1.18}_{1.13}$ &  
15.6 $^{7.45}_{5.16}$ &  
127.3 $^{29.7}_{20.6}$ &  
10.9 \\  
2 & 4.1132 & 
1.26 $^{0.37}_{0.37}$ &  
1.55 $^{0.39}_{0.43}$ &  
6.0e-02 $^{6.23}_{5.9e-02}$ &  
3.8e+03 $^{1.7e+03}_{3.6e+03}$ &  
8.5e+04 \\  
3 & 1.9000 & 
0.27 $^{0.24}_{0.19}$ &  
\nodata $^{ }_{ }$ &  
17.9 $^{7.55}_{10.2}$ &  
120.0 $^{55.0}_{16.8}$ &  
8.9 \\  
4 & 5.3142 & 
1.10 $^{0.55}_{0.48}$ &  
1.97 $^{0.94}_{0.78}$ &  
0.15 $^{8.85}_{0.15}$ &  
4.0e+03 $^{2.0e+03}_{3.8e+03}$ &  
3.5e+04 \\  
5 & 4.0182 & 
5.0e-08 $^{0.25}_{3.6e-08}$ &  
\nodata $^{ }_{ }$ &  
16.7 $^{2.43}_{3.06}$ &  
122.0 $^{8.3}_{7.5}$ &  
9.8 \\  
6 & 1.9934 & 
0.29 $^{0.78}_{0.29}$ &  
0.29 $^{0.78}_{0.29}$ &  
44.3 $^{18.73}_{9.46}$ &  
69.6 $^{7.7}_{24.7}$ &  
2.1 \\  
7 & 4.6242\tablenotemark{*} & 
1.31 $^{0.39}_{0.24}$ &  
1.31 $^{0.39}_{0.24}$ &  
11.5 $^{6.68}_{7.25}$ &  
151.9 $^{79.1}_{31.2}$ &  
17.7 \\  
8 & 4.4812 & 
0.12 $^{0.26}_{0.12}$ &  
0.12 $^{0.47}_{0.12}$ &  
18.0 $^{4.38}_{7.78}$ &  
121.9 $^{31.7}_{11.3}$ &  
9.0 \\  
9 & 3.5878 & 
1.0e-06 $^{0.20}_{1.0e-06}$ &  
\nodata $^{ }_{ }$ &  
4.06 $^{2.35}_{2.33}$ &  
210.9 $^{191.0}_{69.7}$ &  
69.3 \\  
10 & 1.5704 & 
0.82 $^{5.2e-02}_{0.26}$ &  
0.82 $^{5.2e-02}_{0.26}$ &  
4.05 $^{12.02}_{2.60}$ &  
145.5 $^{163.2}_{64.6}$ &  
47.9 \\  
11 & 2.0276 & 
1.4e-06 $^{0.98}_{8.1e-07}$ &  
1.4e-05 $^{1.30}_{1.3e-05}$ &  
12.9 $^{4.42}_{11.9}$ &  
172.2 $^{1.2e+03}_{37.5}$ &  
17.8 \\  
12 & 2.0502 & 
1.30 $^{0.34}_{0.29}$ &  
1.58 $^{0.27}_{0.27}$ &  
5.7e-04 $^{2.52}_{5.6e-04}$ &  
1.6e+03 $^{770.4}_{1.2e+03}$ &  
3.7e+06 \\  
13 & 3.0854 & 
1.6e-06 $^{1.45}_{1.5e-06}$ &  
2.86 $^{1.02}_{1.57}$ &  
10.8 $^{17.65}_{9.40}$ &  
358.8 $^{3.0e+03}_{223.1}$ &  
44.4 \\  
14 & 4.0478 & 
1.45 $^{0.62}_{0.61}$ &  
1.42 $^{1.05}_{1.21}$ &  
30.3 $^{13.11}_{11.7}$ &  
91.5 $^{13.8}_{17.3}$ &  
4.0 \\  
15 & 2.4916 & 
0.82 $^{0.26}_{0.26}$ &  
0.82 $^{0.26}_{0.26}$ &  
39.8 $^{4.02}_{3.65}$ &  
112.7 $^{7.8}_{7.5}$ &  
3.8 \\  
16 & 2.6952 & 
4.5e-07 $^{0.49}_{4.3e-07}$ &  
1.71 $^{1.03}_{0.58}$ &  
17.3 $^{6.35}_{5.46}$ &  
160.1 $^{40.3}_{28.5}$ &  
12.3 \\  
17 & 4.3938 & 
1.0e-07 $^{9.2e-02}_{5.3e-08}$ &  
\nodata $^{ }_{ }$ &  
5.87 $^{5.00}_{3.24}$ &  
140.2 $^{126.4}_{47.0}$ &  
31.8 \\  
18 & 1.5374 & 
0.97 $^{0.21}_{0.27}$ &  
0.97 $^{0.21}_{0.27}$ &  
8.75 $^{4.91}_{5.42}$ &  
84.9 $^{28.2}_{14.6}$ &  
12.9 \\  
19 & 2.8008 & 
0.62 $^{9.6e-02}_{0.10}$ &  
\nodata $^{ }_{ }$ &  
0.17 $^{8.32}_{0.16}$ &  
3.5e+03 $^{1.2e+03}_{3.3e+03}$ &  
2.7e+04 \\  
20 & 3.2332 & 
4.8e-07 $^{5.6e-02}_{4.6e-07}$ &  
\nodata $^{ }_{ }$ &  
11.6 $^{1.10}_{1.11}$ &  
102.5 $^{4.9}_{4.5}$ &  
11.8 \\  
21 & 2.0202 & 
0.11 $^{0.34}_{4.5e-02}$ &  
0.11 $^{0.34}_{4.5e-02}$ &  
18.9 $^{2.17}_{6.60}$ &  
112.5 $^{27.9}_{11.6}$ &  
7.9 \\  
22 & 2.1980 & 
1.2e-06 $^{0.40}_{1.1e-06}$ &  
\nodata $^{ }_{ }$ &  
18.2 $^{3.80}_{4.69}$ &  
211.4 $^{46.6}_{25.8}$ &  
15.5 \\  
23 & 2.4402\tablenotemark{*} & 
6.56 $^{1.07}_{1.45}$ &  
3.46 $^{0.66}_{0.80}$ &  
2.06 $^{4.97}_{2.06}$ &  
2.6e-03 $^{182.0}_{2.6e-03}$ &  
0.0 \\  
24 & 11.8384 & 
0.58 $^{0.28}_{0.27}$ &  
0.58 $^{0.28}_{0.27}$ &  
19.8 $^{2.40}_{2.40}$ &  
127.2 $^{6.3}_{7.0}$ &  
8.6 \\  
25 & 2.5360 & 
1.5e-06 $^{0.13}_{1.4e-06}$ &  
1.5e-06 $^{0.13}_{1.4e-06}$ &  
12.8 $^{1.53}_{1.94}$ &  
216.6 $^{29.7}_{18.2}$ &  
22.5 \\  
26 & 1.0658 & 
2.15 $^{0.15}_{1.29}$ &  
2.15 $^{0.15}_{1.29}$ &  
2.35 $^{10.31}_{1.10}$ &  
267.6 $^{144.0}_{168.2}$ &  
1.5e+02 \\  
27 & 2.0276\tablenotemark{*} & 
0.92 $^{6.3e-02}_{6.4e-02}$ &  
2.35 $^{0.41}_{0.70}$ &  
171.4 $^{0.21}_{81.1}$ &  
8.0e-02 $^{21.6}_{8.0e-02}$ &  
0.0 \\  
28 & 2.5472 & 
7.2e-02 $^{0.18}_{7.2e-02}$ &  
1.22 $^{0.11}_{0.20}$ &  
1.8e-04 $^{5.11}_{1.4e-04}$ &  
3.1e+03 $^{20.7}_{2.8e+03}$ &  
2.3e+07 \\  
29 & 3.9182 & 
2.8e-06 $^{0.37}_{2.7e-06}$ &  
2.8e-06 $^{0.37}_{2.7e-06}$ &  
36.3 $^{4.18}_{6.87}$ &  
157.1 $^{4.1}_{5.5}$ &  
5.8 \\  
30 & 1.7894 & 
5.1e-07 $^{0.63}_{5.1e-07}$ &  
5.1e-07 $^{0.63}_{5.1e-07}$ &  
35.8 $^{2.26}_{5.95}$ &  
127.8 $^{2.8}_{6.6}$ &  
4.8 \\  
31 & 1.6606 & 
1.36 $^{0.31}_{0.52}$ &  
1.36 $^{0.31}_{0.52}$ &  
4.01 $^{6.05}_{2.71}$ &  
248.6 $^{350.8}_{122.2}$ &  
82.7 \\  
32 & 3.0718 & 
1.75 $^{0.21}_{0.34}$ &  
1.75 $^{0.21}_{0.34}$ &  
14.8 $^{19.36}_{9.90}$ &  
101.8 $^{94.2}_{22.4}$ &  
9.2 \\  
33 & 2.2710 & 
1.2e-07 $^{0.23}_{1.2e-07}$ &  
9.8e-02 $^{0.54}_{9.8e-02}$ &  
16.2 $^{2.62}_{4.00}$ &  
167.6 $^{27.4}_{19.6}$ &  
13.8 \\  
34 & 4.2076\tablenotemark{*} & 
1.4e-04 $^{4.42}_{1.4e-04}$ &  
2.34 $^{0.66}_{0.75}$ &  
8.8e-06 $^{1.7e-02}_{8.8e-06}$ &  
1.5e+04 $^{3.8e+03}_{1.2e+04}$ &  
2.3e+09 \\  
35 & 2.2232 & 
1.31 $^{4.5e-02}_{0.12}$ &  
\nodata $^{ }_{ }$ &  
1.86 $^{5.88}_{1.51}$ &  
476.2 $^{775.0}_{318.4}$ &  
3.4e+02 \\  
36 & 2.3106 & 
6.2e-07 $^{0.43}_{2.0e-07}$ &  
6.2e-07 $^{0.43}_{2.0e-07}$ &  
11.3 $^{1.55}_{1.80}$ &  
103.1 $^{6.1}_{6.2}$ &  
12.2 \\  
37 & 2.1400 & 
2.43 $^{1.91}_{0.98}$ &  
4.28 $^{0.86}_{2.85}$ &  
5.37 $^{4.8e+02}_{5.33}$ &  
154.2 $^{152.0}_{153.4}$ &  
38.3 \\  
38 & 1.7024 & 
0.37 $^{0.65}_{0.28}$ &  
0.37 $^{0.65}_{0.28}$ &  
42.3 $^{10.78}_{14.8}$ &  
106.7 $^{6.6}_{9.5}$ &  
3.4 \\  
39 & 3.4314 & 
2.0e-06 $^{0.30}_{1.7e-06}$ &  
2.0e-06 $^{0.30}_{1.7e-06}$ &  
20.1 $^{1.08}_{1.50}$ &  
179.2 $^{2.5}_{7.6}$ &  
11.9 \\  
40 & 1.4178 & 
9.0e-09 $^{0.30}_{9.0e-09}$ &  
\nodata $^{ }_{ }$ &  
69.4 $^{9.54}_{7.85}$ &  
105.5 $^{4.4}_{7.7}$ &  
2.0 \\  
\enddata 
\tablenotetext{a}{R$_s$ is not a fitted parameter.
It has been calculated using the best-fit $v_{200}$ and $c$ 
values to facilitate comparison with the PI scale length 
parameter.} 
\tablenotetext{*}{Indicates that the halo fit is 
no better than the stars-only fit at the 95\% level.} 
\end{deluxetable} 

\begin{deluxetable}{lccccc} 
\tablewidth{0pt} 
\tablecaption{Best-fit $b^2$ values, parameter 
values, and uncertainties for POWER models.\label{ptab}} 
\tablehead{\colhead{Galaxy} & 
\colhead{$b^2$} & 
\colhead{\mld (M$_{\sun}$/L$_{\sun}$)} & 
\colhead{\mlb (M$_{\sun}$/L$_{\sun}$)} & 
\colhead{$v_0$ (km/s)} & 
\colhead{$\alpha$}} 
\startdata 
1 & 1.4170 & 
9.3e-08 $^{0.16}_{1.6e-16}$ &  
2.69 $^{0.92}_{0.83}$ &  
158.2 $^{5.33}_{8.50}$ &  
1.2 $^{0.1}_{0.2}$ \\  
2 & 4.1074 & 
1.47 $^{0.50}_{0.45}$ &  
1.79 $^{0.41}_{0.53}$ &  
162.5 $^{30.0}_{35.8}$ &  
0.7 $^{0.4}_{0.7}$ \\  
3 & 1.9188 & 
0.56 $^{0.19}_{0.14}$ &  
\nodata $^{ }_{ }$ &  
134.2 $^{19.9}_{25.5}$ &  
1.2 $^{0.1}_{0.4}$ \\  
4 & 5.3118 & 
1.56 $^{0.25}_{0.32}$ &  
1.56 $^{0.25}_{0.32}$ &  
161.8 $^{18.2}_{18.9}$ &  
0.8 $^{0.3}_{0.5}$ \\  
5 & 4.0946 & 
0.89 $^{0.22}_{0.21}$ &  
\nodata $^{ }_{ }$ &  
151.4 $^{9.57}_{11.0}$ &  
1.3 $^{0.1}_{0.2}$ \\  
6 & 2.0340\tablenotemark{*} & 
1.41 $^{0.27}_{0.36}$ &  
2.46 $^{0.76}_{0.46}$ &  
48.7 $^{36.5}_{48.7}$ &  
1.6 $^{3.5}_{1.6}$ \\  
7 & 4.6356\tablenotemark{*} & 
1.75 $^{0.22}_{0.22}$ &  
1.75 $^{0.22}_{0.22}$ &  
140.7 $^{23.3}_{27.4}$ &  
1.1 $^{0.2}_{0.5}$ \\  
8 & 4.4868 & 
0.43 $^{0.14}_{0.13}$ &  
0.43 $^{0.14}_{0.13}$ &  
156.5 $^{9.76}_{9.43}$ &  
1.4 $^{7.1e-02}_{9.7e-02}$ \\  
9 & 3.5878 & 
0.39 $^{0.21}_{0.22}$ &  
\nodata $^{ }_{ }$ &  
117.1 $^{10.3}_{11.1}$ &  
0.8 $^{0.2}_{0.2}$ \\  
10 & 1.5710 & 
0.83 $^{8.2e-02}_{0.15}$ &  
0.83 $^{8.2e-02}_{0.15}$ &  
111.8 $^{8.24}_{7.84}$ &  
1.2 $^{0.3}_{0.3}$ \\  
11 & 2.0196 & 
9.3e-02 $^{1.29}_{9.3e-02}$ &  
9.3e-02 $^{1.29}_{9.3e-02}$ &  
175.7 $^{9.51}_{61.4}$ &  
1.2 $^{5.7e-02}_{1.2}$ \\  
12 & 2.0456 & 
1.67 $^{0.19}_{0.46}$ &  
1.63 $^{0.30}_{0.24}$ &  
105.5 $^{22.5}_{19.4}$ &  
1.4e-05 $^{1.0}_{1.4e-05}$ \\  
13 & 3.0854 & 
4.7e-06 $^{2.40}_{4.7e-06}$ &  
2.64 $^{2.12}_{1.37}$ &  
216.9 $^{7.92}_{87.8}$ &  
1.1 $^{0.2}_{1.1}$ \\  
14 & 4.0674\tablenotemark{*} & 
2.73 $^{0.11}_{0.28}$ &  
3.01 $^{0.64}_{0.63}$ &  
45.5 $^{31.9}_{38.6}$ &  
1.4e-07 $^{2.0}_{1.4e-07}$ \\  
15 & 2.6250 & 
1.84 $^{0.16}_{0.19}$ &  
1.84 $^{0.16}_{0.19}$ &  
142.2 $^{15.2}_{14.3}$ &  
1.9 $^{5.5e-02}_{5.4e-02}$ \\  
16 & 2.7340 & 
2.5e-06 $^{0.59}_{1.9e-06}$ &  
1.90 $^{1.11}_{0.97}$ &  
209.2 $^{4.99}_{34.2}$ &  
1.4 $^{0.2}_{0.1}$ \\  
17 & 4.4164 & 
7.8e-08 $^{0.18}_{2.0e-08}$ &  
\nodata $^{ }_{ }$ &  
123.2 $^{7.58}_{9.72}$ &  
1.1 $^{0.2}_{0.2}$ \\  
18 & 1.5448 & 
1.33 $^{8.9e-02}_{9.9e-02}$ &  
1.33 $^{8.9e-02}_{9.9e-02}$ &  
71.2 $^{11.8}_{12.7}$ &  
1.0 $^{0.3}_{0.4}$ \\  
19 & 2.8006 & 
1.07 $^{6.0e-02}_{0.63}$ &  
\nodata $^{ }_{ }$ &  
138.7 $^{27.7}_{9.91}$ &  
4.1e-06 $^{1.2}_{2.0e-06}$ \\  
20 & 3.2848 & 
0.78 $^{0.29}_{0.25}$ &  
\nodata $^{ }_{ }$ &  
102.9 $^{10.1}_{13.8}$ &  
1.3 $^{0.1}_{0.2}$ \\  
21 & 2.0262 & 
0.78 $^{0.77}_{0.35}$ &  
1.35 $^{1.65}_{1.13}$ &  
139.0 $^{13.8}_{20.0}$ &  
1.1 $^{0.2}_{1.1}$ \\  
22 & 2.2136 & 
0.31 $^{0.89}_{0.31}$ &  
\nodata $^{ }_{ }$ &  
278.0 $^{9.37}_{15.4}$ &  
1.3 $^{5.8e-02}_{0.2}$ \\  
23 & 2.4402\tablenotemark{*} & 
6.56 $^{0.92}_{1.19}$ &  
3.46 $^{0.72}_{0.67}$ &  
1.0e-02 $^{50.8}_{3.0e-10}$ &  
1.1 $^{5.0}_{1.1}$ \\  
24 & 11.8892\tablenotemark{*} & 
2.25 $^{0.22}_{0.23}$ &  
2.25 $^{0.22}_{0.23}$ &  
76.2 $^{26.6}_{39.2}$ &  
1.5 $^{0.1}_{0.3}$ \\  
25 & 2.5674 & 
0.53 $^{0.34}_{0.33}$ &  
0.53 $^{0.34}_{0.33}$ &  
213.5 $^{11.3}_{11.4}$ &  
1.1 $^{0.1}_{0.1}$ \\  
26 & 1.0642 & 
2.56 $^{0.41}_{0.42}$ &  
2.56 $^{0.41}_{0.42}$ &  
94.8 $^{12.8}_{12.0}$ &  
0.7 $^{0.3}_{0.6}$ \\  
27 & 2.0276\tablenotemark{*} & 
0.92 $^{6.8e-02}_{6.9e-02}$ &  
2.35 $^{0.57}_{0.61}$ &  
2.1e-02 $^{28.3}_{2.1e-02}$ &  
2.0 $^{2.3}_{2.0}$ \\  
28 & 2.5464 & 
0.36 $^{0.52}_{0.36}$ &  
1.64 $^{0.47}_{0.56}$ &  
155.1 $^{18.6}_{26.3}$ &  
0.7 $^{0.3}_{0.7}$ \\  
29 & 4.0170 & 
1.42 $^{0.23}_{0.21}$ &  
1.42 $^{0.23}_{0.21}$ &  
208.5 $^{13.5}_{15.0}$ &  
1.3 $^{0.1}_{0.2}$ \\  
30 & 1.8488 & 
2.33 $^{0.26}_{0.31}$ &  
2.33 $^{0.26}_{0.31}$ &  
140.4 $^{17.9}_{18.7}$ &  
1.3 $^{0.2}_{0.2}$ \\  
31 & 1.6564 & 
1.86 $^{0.38}_{0.36}$ &  
1.86 $^{0.38}_{0.36}$ &  
128.9 $^{11.1}_{12.5}$ &  
0.8 $^{0.3}_{0.4}$ \\  
32 & 3.0768 & 
1.86 $^{0.26}_{0.32}$ &  
1.86 $^{0.26}_{0.32}$ &  
133.4 $^{14.1}_{15.8}$ &  
1.4 $^{0.3}_{0.5}$ \\  
33 & 2.2718 & 
7.3e-02 $^{0.48}_{7.3e-02}$ &  
7.3e-02 $^{0.48}_{7.3e-02}$ &  
215.8 $^{4.49}_{21.3}$ &  
1.4 $^{3.5e-02}_{7.4e-02}$ \\  
34 & 4.2046 & 
2.0e-05 $^{3.80}_{1.8e-05}$ &  
3.02 $^{1.25}_{1.60}$ &  
305.5 $^{18.7}_{132.6}$ &  
0.9 $^{0.3}_{0.4}$ \\  
35 & 2.2232 & 
1.34 $^{0.12}_{0.18}$ &  
\nodata $^{ }_{ }$ &  
157.9 $^{10.8}_{11.2}$ &  
1.0 $^{0.3}_{0.3}$ \\  
36 & 2.3404 & 
1.95 $^{0.44}_{0.44}$ &  
1.95 $^{0.44}_{0.44}$ &  
108.0 $^{11.8}_{12.6}$ &  
1.1 $^{0.1}_{0.2}$ \\  
37 & 2.1290 & 
2.21 $^{1.29}_{1.03}$ &  
3.36 $^{1.72}_{1.10}$ &  
118.3 $^{30.2}_{56.3}$ &  
1.6 $^{0.3}_{1.6}$ \\  
38 & 1.7194 & 
1.55 $^{0.28}_{0.30}$ &  
1.55 $^{0.28}_{0.30}$ &  
142.2 $^{17.0}_{20.3}$ &  
1.5 $^{0.2}_{0.3}$ \\  
39 & 3.5774 & 
0.60 $^{0.53}_{0.42}$ &  
0.60 $^{0.53}_{0.42}$ &  
241.9 $^{14.7}_{20.8}$ &  
1.6 $^{2.6e-02}_{3.2e-02}$ \\  
40 & 1.6092 & 
1.35 $^{0.51}_{0.37}$ &  
\nodata $^{ }_{ }$ &  
147.8 $^{34.4}_{50.4}$ &  
1.8 $^{0.1}_{0.1}$ \\  
\enddata 
\tablenotetext{*}{Indicates that the halo fit is 
no better than the stars-only fit at the 95\% level.} 
\end{deluxetable} 

\end{document}